\documentclass[sigconf]{acmart}
\AtBeginDocument{%
  }

\usepackage{algorithm}
\usepackage{algorithmic}
\usepackage{color}
\usepackage[table]{xcolor}
\usepackage{enumitem}
\definecolor{myblue}{RGB}{221,235,248}
\definecolor{myorange}{RGB}{251,229,214}
\definecolor{mygreen}{RGB}{213,232,213}

\usepackage{booktabs}       
\usepackage{amsfonts}       
\usepackage{nicefrac}       
\usepackage{microtype}      
\usepackage{xcolor}         
\usepackage{subcaption} 
\usepackage{amsmath} 
\usepackage{newfloat}

\copyrightyear{2026}
\acmYear{2026}
\setcopyright{cc}
\setcctype{by}
\acmConference[CODASPY '26] {Proceedings of the Sixteenth ACM Conference on Data and Application Security and Privacy}{June 23--25, 2026}{Frankfurt am Main, Germany.}
\acmBooktitle{Proceedings of the Sixteenth ACM Conference on Data and Application Security and Privacy (CODASPY '26), June 23--25, 2026, Frankfurt am Main, Germany}
\acmDOI{10.1145/3800506.3803494}
\acmISBN{979-8-4007-2562-3/2026/06}

\begin{document}

\title{Measuring the Robustness of Audio Deepfake Detection under Real-World Corruption}


\author{Xiang Li}
\email{xl5@fordham.edu}
\orcid{0009-0007-7467-6221}
\affiliation{%
  \institution{Fordam University}
  \city{New York City}
  \state{New York}
  \country{USA}
}

\author{Pin-Yu Chen}
\email{pin-yu.chen@ibm.com}
\orcid{0000-0003-1039-8369}
\affiliation{%
  \institution{IBM Research}
  \city{Yorktown Heights}
  \state{New York}
  \country{USA}}

\author{Wenqi Wei}
\email{wwei23@fordham.edu}
\orcid{0000-0001-9177-114X}
\affiliation{%
  \institution{Fordham University}
  \city{New York City}
  \state{New York}
  \country{USA}
}

\renewcommand{\shortauthors}{Xiang Li, Pin-Yu Chen, \& Wenqi Wei}

\begin{abstract}
Deepfakes have emerged as a widespread and rapidly escalating concern in generative AI, spanning various media types such as images, audio, and videos. Among these, audio deepfakes are particularly alarming due to the growing accessibility of high-quality voice synthesis tools and the ease with which synthetic speech can be distributed via platforms like social media and robocalls. Consequently, detecting audio deepfakes is critical in combating the misuse of AI-synthesized speech. However, real-world audio is often subject to various corruptions, such as noise, modification, and compression, that may significantly impact detection performance. In this work, we systematically evaluate the robustness of 10 audio deepfake detection models against 18 common corruption types, grouped into categories: noise perturbation, audio modification, and compression. Using both traditional deep learning models and state-of-the-art foundation models, our study yields four key insights. (1) Most models demonstrate strong robustness to noise but they are notably more vulnerable to audio modifications and compression, especially when neural codecs are applied. (2) Speech foundation models generally outperform traditional models across most corruption scenarios, likely due to their extensive pre-training on large-scale and diverse audio datasets. (3) Increasing model size improves robustness, though with diminishing returns. (4) Robustness to unseen corruptions can be enhanced by targeted data augmentation during training or by applying speech enhancement techniques at inference time. 
These findings highlight the importance of comprehensive evaluation against diverse corruption types and developing more robust audio deepfake detection frameworks to ensure reliability in practical deployment settings. We further advocate that future research in deepfake detection across all media formats should account for the diverse and often unpredictable distortions common in real-world environments.
\end{abstract}

\ccsdesc[500]{Security and privacy~Usability in security and privacy}
\ccsdesc[500]{General and reference~Evaluation}
\ccsdesc[500]{General and reference~Reliability}

\keywords{Deepfake Detection; Robustness; Security}


\maketitle

\section{Introduction}
\label{sec:intro}

Deepfakes have become a growing concern of generative Artificial Intelligence (AI) across multiple media types such as images, audio, and videos. Among these, audio deepfakes pose particularly severe risks due to their ability to mimic human speech with high fidelity. Recent advances in Text-to-Speech (TTS) and Voice-Conversion (VC) technologies have enabled the efficient generation of high-quality and
realistic human-like audio~\cite{vyas2023audiobox,wang2023neural}. While these developments have unlocked valuable applications in education, accessibility, and language preservation, they also present serious ethical and societal risks stemming from their potential misuse. These dangers are not just hypothetical or confined to controlled environments. They are unfolding in high-stakes, real-world scenarios\footnote{\url{ww.wired.com/story/slovakias-election-deepfakes-show-ai-is-a-danger-to-democracy/}}\footnote{\url{https://www.cnn.com/2024/01/22/politics/fake-joe-biden-robocall/index.html}}. 
Recent studies reinforce the severity of these threats: \cite{groh2024human} showed that humans struggle to differentiate real political speeches from those generated by state-of-the-art TTS models. Likewise, \cite{cooke2024good} found that people 
are no better than chance at identifying fake media-including audio, images, and video. 
Three key factors drive the misuse potential of AI-generated speech: (1) the ease with which AI-synthesized speech can be created and distributed via platforms like social media and robocalls; (2) the high quality of synthesized audio and voice cloning, making it increasingly difficult to distinguish from genuine human speech; and (3) the inherent human trust in voice and audiovisual content, allowing even unsophisticated deepfakes or synthetic media effective tools for misinformation. These evidences reveal not only the deceptive power of synthetic audio, but also the growing vulnerability of the public to AI-driven deception and manipulation. 

\begin{figure*}[!t]
\centering
\includegraphics[width=0.7\textwidth]{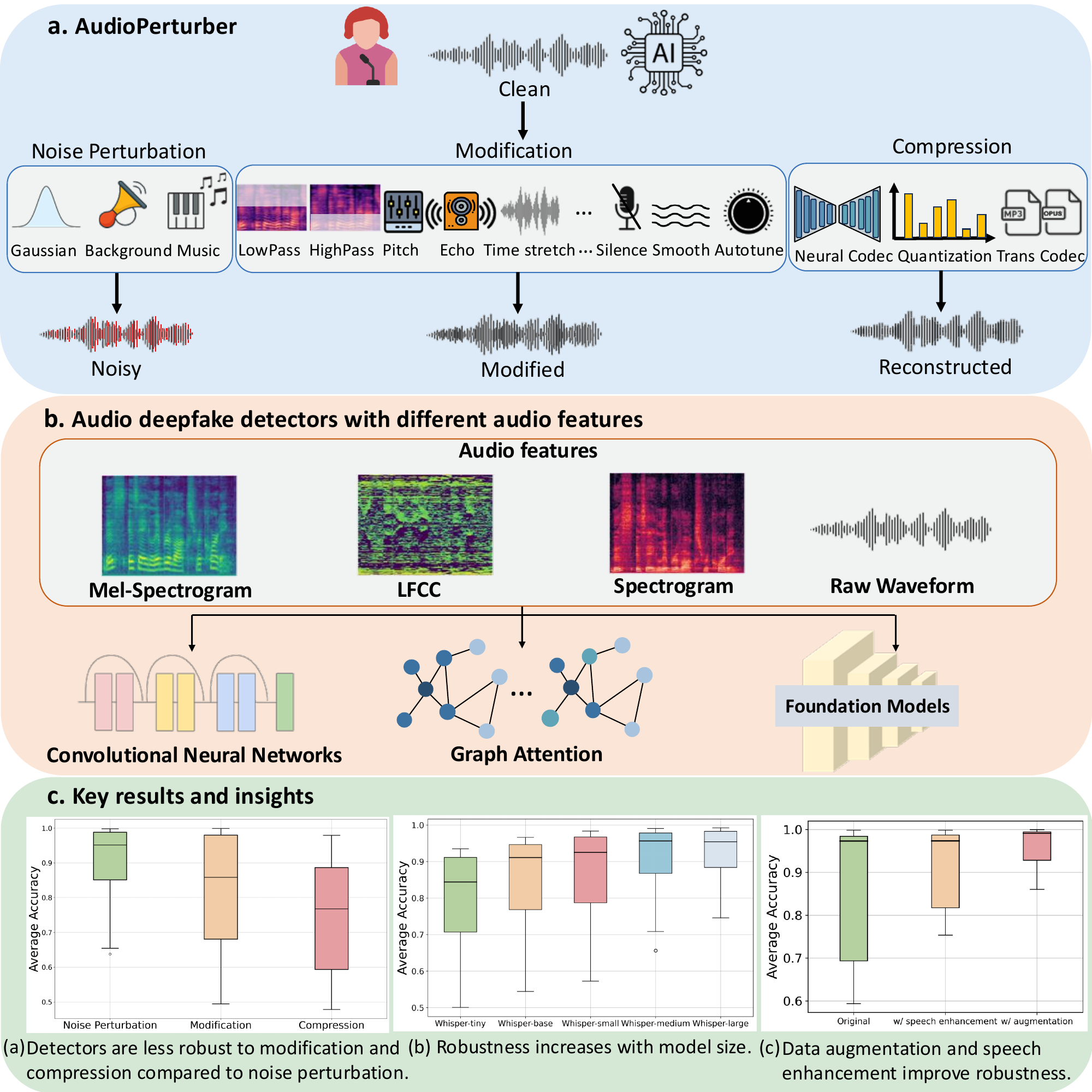}
\caption{Overview of the proposed framework. (a)  \textit{AudioPerturber} covers 18 audio corruption techniques across three categories: noise perturbation, modification, and compression (including 4 types of neural codecs and 2 types of transmission codecs). (b) We evaluate 10 state-of-the-art audio deepfake detection models using diverse audio features and architectures, such as convolutional networks, graph attention mechanisms, and foundation models. (c) The evaluation shows that
while these models are generally robust to noise perturbations, they are significantly more vulnerable to audio modifications and compression artifacts. Larger model sizes tend to improve robustness, and incorporating data augmentation and speech enhancement techniques can further boost models' robustness under common audio corruptions.}
\label{fig:overview}
\vspace{-0.3cm}
\end{figure*}

In response to the growing threat of audio deepfakes, an expanding body of research~\cite{tak2021endrawnet,lavrentyeva2019stc,jung2022aasist,tak2021end,wang2021investigating,tak2022automatic,kawa2023improved} has focused on developing audio deepfake detection models capable of distinguishing synthetic from real speech. While these models have shown strong performance in controlled settings, they often fail to generalize beyond the specific TTS systems or data distributions on which they were trained~\cite{muller2022does,li2024sonar}. A key challenge for real-world deployment is robustness. In practice, audio is frequently degraded by various perturbations that can compromise detector reliability. Common acoustic corruptions, such as Gaussian noise and environmental background noise (e.g., street sounds, chatter, or wind) continue to pose unresolved challenges. Moreover, audio signals often undergo intentional or unintentional modifications during transmission, editing, or playback. For example, standard audio processing techniques like low-pass and high-pass filtering, intended to improve clarity or suppress unwanted frequencies, can unintentionally distort features essential for detection. 
Likewise, pitch-shifting and time-stretching are widely used in media production and audio personalization by modifying the frequency or timing of the audio, further complicating the task. Furthermore, streaming media has surged in global internet usage, with audio and video accounting for approximately 82\% of internet traffic in 2021~\cite{cisco2021global}. To manage growing traffic, audio compression techniques like MP3 and Opus~\cite{valin2012definition} are frequently utilized, along with emerging neural codecs~\cite{zeghidour2021soundstream,defossez2022high,wu2023audiodec,ju2024naturalspeech,kumar2024high}, which introduce additional distortions that can hinder detection accuracy. However, there remains a lack of systematic evaluation of deepfake detectors under real-world audio corruptions. Understanding how these detectors perform under audio corruption is essential for assessing their limitations and determining their readiness for deployment in practical, high-stakes environments. 
\begin{table*}[!ht]
\centering
\caption{Details of the audio perturbation types included in \textit{AudioPerturber}. \colorbox{myblue!50}{Blue}, \colorbox{myorange!50}{Orange}, and \colorbox{mygreen!50}{Green} denote noise perturbation, modifications, and compression, respectively.}
\label{tab:perturbation_details}
\resizebox{0.78\textwidth}{!}{
\begin{tabular}{c|c|c|c}
\toprule
\textbf{Perturbation} & \textbf{Parameter $\mathcal{P}$} & \textbf{Range of $\mathcal{P}$} & \textbf{Description} \\
\midrule

\rowcolor{myblue!50}Gaussian Noise & SNR (dB) & [5, 40] & Adds Gaussian white noise at the specified SNR  \\
\hline
\rowcolor{myblue!50}Background Noise & SNR (dB) & [5, 40] & Mixes environmental background noise into the audio \\
\hline
\rowcolor{myblue!50}Background Music & SNR (dB) & [5, 40] & Overlays background music at the given SNR \\
\hline
\rowcolor{myorange!50}Pitch Shift & Semitone & [-2, 2] & Alters the pitch by a given semitones without changing the duration \\
\hline
\rowcolor{myorange!50}Autotune & Key & [C,D,..., B] & \begin{tabular}[c]{@{}c@{}} Adjusts pitch to the closest note within a specified\\ musical scale (e.g., C major)\end{tabular}  \\
\hline
\rowcolor{myorange!50}Echo & Delay (sec) & [0.1, 0.9] & Adds a delayed and decayed echo effect \\
\hline
\rowcolor{myorange!50}Time Stretch & Speed Factor & [0.7, 1.5] & Changes playback speed of the audio \\
\hline
\rowcolor{myorange!50}Highpass Filter & Cutoff ratio & [0.1,0.5] & Removes frequencies below the cutoff threshold \\
\hline
\rowcolor{myorange!50}Lowpass Filter & Cutoff ratio & [0.1,0.5] & Removes frequencies above the cutoff threshold \\
\hline
\rowcolor{myorange!50}Smooth & Window size &  [6, 26] & Applies 1-D Gaussian smoothing to the waveform \\
\hline
\rowcolor{myorange!50}Silence Insertion & Length ratio & [0.1, 0.5] & \begin{tabular}[c]{@{}c@{}} Replaces a randomly selected segment of length\\ $\mathcal{P}\times$ audio\_length with silence\end{tabular}  \\  
\hline
\rowcolor{mygreen!50}Opus & Bitrate (kbps) & [16, 256] & Commonly adopted audio codec \\
\hline
\rowcolor{mygreen!50}MP3 Compression & Bitrate (kbps) & [8, 40] & Commonly adopted audio codec \\
\hline
\rowcolor{mygreen!50}Quantization & Bit levels & [2,10] & Reduces bit depth to $n$-bit discrete levels \\
\hline
\rowcolor{mygreen!50}EnCodec & Bandwidth (kHz) & [1.5, 24.0] & Neural network-based audio codec \\
\hline
\rowcolor{mygreen!50}AudioDec & Bitrate (kbps) & default & Neural network-based audio codec \\
\hline
\rowcolor{mygreen!50}FACodec & Bitrate (kbps) & default & Neural network-based audio codec \\
\hline
\rowcolor{mygreen!50}DAC & Bitrate (kbps) & default & Neural network-based audio codec \\
\bottomrule

\end{tabular}}
\vspace{-0.2cm}
\end{table*}

To bridge this gap, this work focuses on systematically evaluating the robustness of audio deepfake detection models under real-world audio corruptions. By examining how these systems perform when exposed to corrupted or intentionally manipulated inputs, our study contributes to the broader goal of building trustworthy and resilient AI systems and ongoing efforts in AI alignment by addressing challenges in robustness, security, and safety evaluation. Our main contributions are as follows:

\begin{itemize}
    \item We introduce \textit{AudioPerturber}, a comprehensive evaluation framework encompassing 18 common real-world audio corruptions grouped into three categories: noise perturbation, audio modification, and compression. Using this framework, we systematically assess the robustness of 10 state-of-the-art audio deepfake detectors and present several key observations. 
    
    \item We conduct a case study on the Whisper model family~\cite{radford2023robust} to examine whether increasing model size enhances robustness in audio deepfake detection against audio corruption. Our findings suggest that larger models tend to exhibit improved robustness, likely due to their ability to learn more invariant and robust feature representations.
    
    \item We further explore two practical strategies to improve detection robustness: (1) incorporating corruption-based data augmentation during training or fine-tuning, and (2) applying speech enhancement methods at inference time. Both approaches show measurable gains in detection performance under corrupted audio conditions.
\end{itemize}

\section{Related Work}
\label{sec:related_work}

Robustness under real-world corruptions has been widely studied in other domains. In image classification, benchmarks such as ImageNet-C~\cite{hendrycks2019benchmarking} show that models with high clean accuracy can degrade significantly under noise, blur, and compression. In 3D point cloud recognition, models are sensitive to occlusion, point dropout, and sensor noise common in LiDAR data~\cite{sun2022benchmarking}. In watermark detection, robustness to compression, scaling, and re-encoding is critical due to repeated media processing~\cite{liu2024audiomarkbench}. These works underscore that clean-test performance often overestimates real-world reliability. 

Similarly, audio data is widely shared, easily transmitted, and easily manipulated, facing diverse post-processing effects, which motivates corruption-based robustness evaluation in audio deepfake detection. Existing datasets such as ASVspoof 2021~\cite{liu2023asvspoof} and VoiceWukong~\cite{yan2024voicewukong} incorporate a narrow set of perturbations, including noise injection, reverberation, and standard codec compression. CD-ADD~\cite{li2024cross} further includes low-pass filtering and Encodec-based compression. The In-the-Wild dataset~\cite{muller2022does} collects both real and synthetic audio from online sources, better reflecting real-world conditions. However, these datasets are limited in both the diversity and severity of perturbations considered, and only evaluate a small set of detection models, thus not fully revealing the vulnerabilities of current detection systems.

To address this gap, we introduce a comprehensive and extensible evaluation framework that captures a broad range of common audio corruptions across varying severity levels. Our study specifically targets naturally occurring corruptions — noise, compression, and channel effects — that are highly prevalent in real-world audio, complementing prior work on adversarial robustness~\cite{uddin2025adversarial}. Our tool, AudioPerturber, supports flexible and modular application of 18 distinct perturbation types. Leveraging this framework, we systematically evaluate the robustness of 10 different models—including both dedicated audio deepfake detectors and general-purpose speech foundation models—under diverse corruption scenarios.

\section{Audio Perturbation and Evaluation Methods}
\label{sec:perturb}

\subsection{AudioPerturber}
To simulate real-world corruptions that audio deepfake detectors may encounter in practical applications, we introduce \textbf{AudioPerturber} as shown in \textbf{Figure~\ref{fig:overview}}. AudioPerturber is a modular and extensible framework designed to systematically apply a wide range of audio corruptions. AudioPerturber covers 18 perturbation types across three categories:

\begin{itemize}
    \item \textbf{Noise Perturbation:} adding external audio sources such as Gaussian noise, environmental background sound~\citep{richey2018voices}, and background music~\citep{agostinelli2023musiclm}, with varying signal-to-noise ratios (SNR).
    \item \textbf{Audio Modification:} altering the audio signal through a variety of transformations, including pitch shifting, autotuning, time stretching, echo, silence insertion, (high-pass and low-pass) frequency filtering, and Gaussian smoothing.
    \item \textbf{Compression:} applying both traditional transmission codecs (e.g., MP3 and Opus~\citep{valin2012definition}) and advanced neural audio codecs (e.g., EnCodec~\citep{defossez2022high}, DAC~\cite{kumar2024high}, FACodec~\cite{ju2024naturalspeech}, and AudioDec~\cite{wu2023audiodec}). 
\end{itemize}

Each perturbation is parameterized by a control variable $\mathcal{P}$, allowing us to simulate a range of corruption severity, from mild to severe. The detailed descriptions, parameters, and severity settings are summarized in \textbf{Table~\ref{tab:perturbation_details}}. 

\begin{table}[t]
\caption{Accuracy, AUROC, and EER(\%) on the clean test set of Wavefake dataset.}
\label{tab:clean_results}
\centering
\resizebox{0.42\textwidth}{!}{
\begin{tabular}{c|ccc}
\toprule
\textbf{Model} & \textbf{Accuracy} ($\uparrow$) & \textbf{AUROC}($\uparrow$) & \textbf{EER(\%)($\downarrow$)} \\ 
\hline
LFCC-LCNN & 0.9984 & 0.9999 & 0.153  \\ 
Res\_Spec. & 0.9924 & 0.9924 & 0.076  \\ 
RawNet2 & 0.9416 & 0.9592 & 5.839  \\ 
RawGATST & 0.9988 & 0.9999 & 0.115  \\ 
AASIST & 0.9992 & 0.9999 & 0.076  \\
CLAP & 0.9996 & 0.9999 & 0.038  \\ 
Whisper & 0.9935 & 0.9997 & 0.649  \\ 

Wave2Vec2 & 0.9874 & 0.9987 & 1.259   \\ 
HuBERT & 0.9931 & 0.9996 & 0.687 \\
Wave2Vec2BERT & 0.9996 & 0.9999 & 0.038   \\ 
\bottomrule
\end{tabular}}
 \vspace{-0.3cm}
\end{table}

\subsection{Evaluation settings}
\textbf{Dataset}. We use the Wavefake dataset~\cite{frank2021wavefake}, which contains approximately 196 hours of synthetic audio generated using six vocoder architectures. 
All synthetic samples are derived from the LJSPEECH dataset~\cite{ljspeech17}. The dataset is split into training, validation, and test sets with a ratio of 0.7:0.1:0.2. As shown in \textbf{Table~\ref{tab:clean_results}}, \textbf{all detection models achieve near-perfect performance. Our goal here is to examine even under such favorable conditions how detection performance changes when realistic corruptions are introduced. Therefore, we focus on a single dataset to isolate corruption effects.}

\textbf{Detection models}. We evaluate 10 audio deepfake detection models from~\cite{li2024sonar}. These models include both traditional deepfake detectors. The traditional models, including LFCC-LCNN~\cite{lavrentyeva2019stc}, ResNet\_Spec.~\cite{zang2024singfake}, RawNet2~\cite{tak2021endrawnet}, AASIST~\cite{jung2022aasist}, RawGATST~\cite{tak2021end} are specifically designed for synthetic speech detection. The foundation models, including CLAP~\cite{elizalde2023clap}, Whisper~\cite{radford2023robust}, Wave2Vec2~\cite{baevski2020wav2vec}, HuBERT~\cite{hsu2021hubert}, and Wave2Vec2BERT~\cite{barrault2023seamless}, are not trained for deepfake detection but serve as robust feature extractors due to their pretraining on large-scale audio corpus.

\textbf{Evaluation Metrics}. To comprehensively assess the performance of audio deepfake detectors, we consider the following metrics: (1) \textit{Equal Error Rate} (EER) is defined as the point on the Receiver Operating Characteristic (ROC) curve where the False Positive Rate (FPR) equals the False Negative Rate (FNR), with lower values indicating better detection performance; (2) \textit{Accuracy} measures the overall correctness of the detection model’s predictions and is reported using the same decision threshold as the EER point to ensure consistency; (3) \textit{AUROC} measures the area under the ROC curve, providing a threshold-independent evaluation of a model’s ability to distinguish between real and fake audio. 

\textbf{Perturbation Quality.}
To assess the quality of the corrupted audio, we employ two standard metrics: Signal-to-Noise Ratio (SNR) and ViSQOL \cite{hines2015visqol}. SNR measures the ratio between signal power and noise power relative to the clean reference, with higher SNR indicating better audio quality. ViSQOL models human perception of audio quality, assigning a score from 1 to 5, where higher scores correspond to better preservation of the reference audio. A ViSQOL score of 3 or above typically represents acceptable audio quality. We primarily rely on ViSQOL, as it more reliably captures perceptual degradation than SNR~\cite{hines2015visqol}. In our experiments, we focus on corruption levels that maintain ViSQOL $\ge$ 3 to ensure the audio remains intelligible and realistic, conditions that better reflect real-world use cases. Audio that is severely degraded to the point of being unrecognizable is excluded, as it is unlikely to occur or be useful in practice. Unless otherwise specified, this quality criterion is applied throughout our evaluation.

\begin{figure*}[!t]
    \centering

    \begin{subfigure}{0.75\textwidth}
        \centering
        \includegraphics[width=\textwidth]{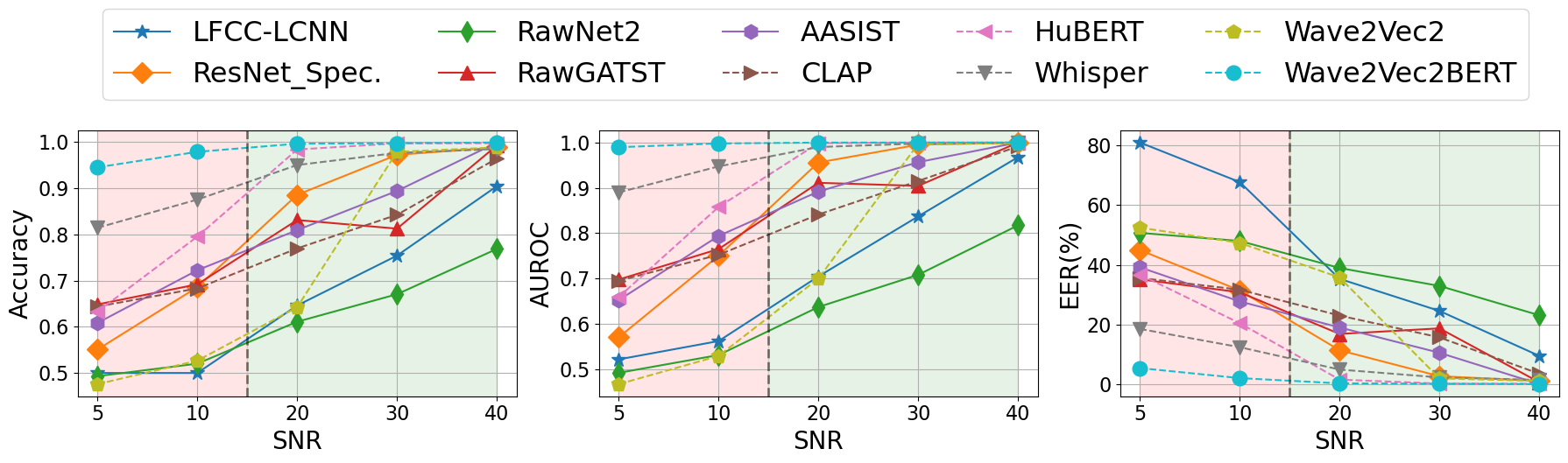}
        \caption{Gaussian noise}
        \label{fig:gaussian_noise}
    \end{subfigure}

    \begin{subfigure}{0.75\textwidth}
        \centering
        \includegraphics[width=\textwidth]{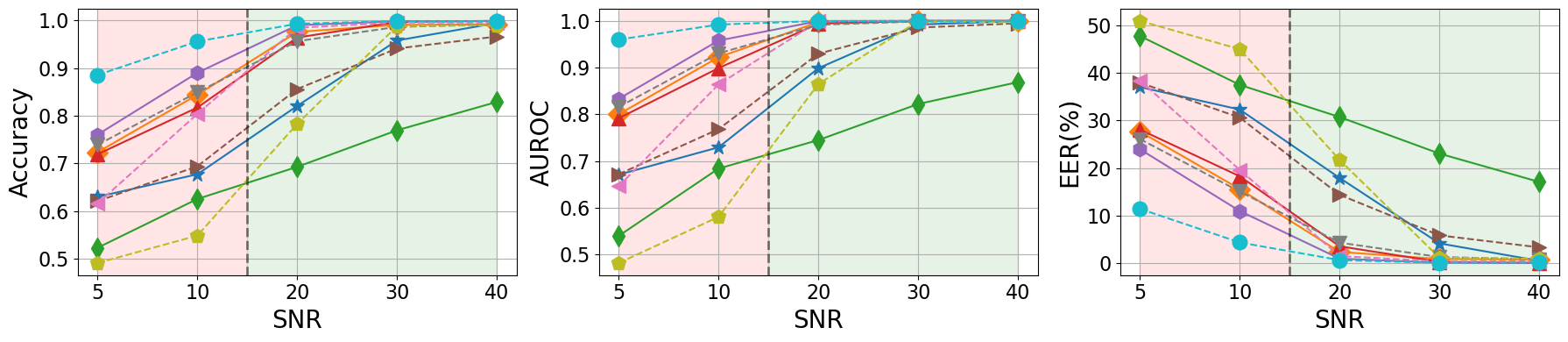}
        \caption{Background noise}
        \label{fig:bg_noise}
    \end{subfigure}

    \begin{subfigure}{0.75\textwidth}
        \centering
        \includegraphics[width=\textwidth]{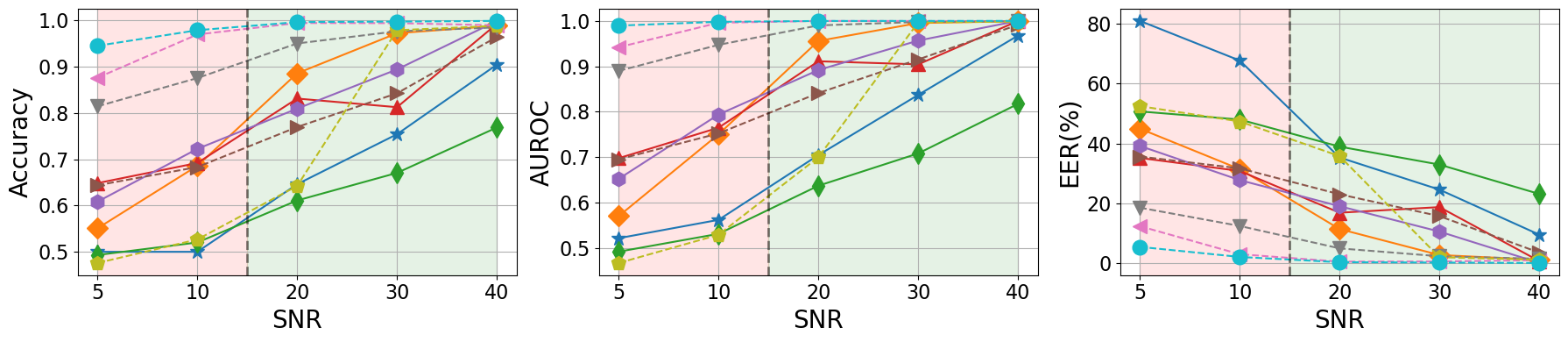}
        \caption{Background Music}
        \label{fig:music}
    \end{subfigure}

\caption{Robustness to noise perturbation across varying SNRs. Green-shaded regions represent SNR levels where audio quality remains acceptable ($\text{ViSQOL} \geq 3$). Detailed ViSQOL scores for each corruption type and severity level are provided in Appendix~\ref{appendix:audio_quality}. 
}
\label{fig:noise_perturbation}
\end{figure*}
\section{Evaluation}
\label{sec:results}

In this section, we empirically investigate our research questions regarding the robustness of audio deepfake detectors under real-world corruptions. Specifically, we aim to answer: 
\begin{enumerate}
    \item whether different detectors exhibit consistent robustness across various corruption types?
    \item how perceptual audio quality relates to detection performance?
    \item whether large-scale pretrained foundation models demonstrate improved corruption invariance?
    \item what training- or inference-time strategies can effectively mitigate performance degradation?
\end{enumerate}

To this end, we analyze the impact of multiple categories of audio corruptions, including noise perturbations, audio modifications, and compression artifacts. We first evaluate how these corruption types affect detection performance across different models. We then study the role of model scale and pretraining in improving robustness, followed by an investigation of mitigation strategies such as data augmentation and speech enhancement.

\begin{figure*}[!t]
    \centering
        \begin{subfigure}{0.75\textwidth}
        \centering
        \includegraphics[width=\textwidth]{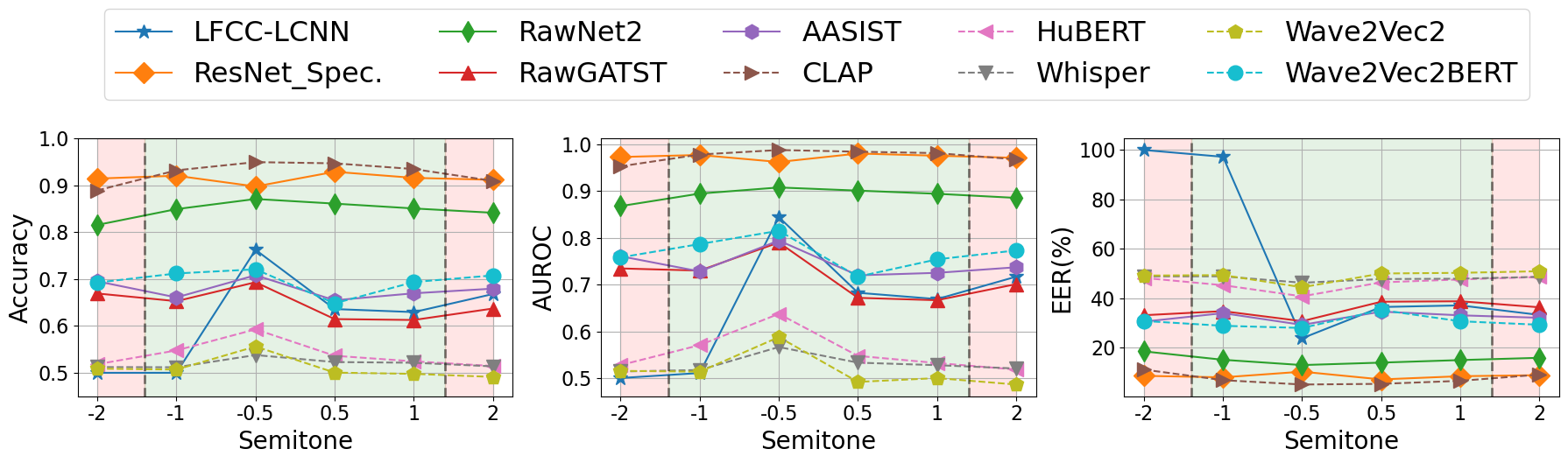}
        \caption{Pitch shifting}
        \label{fig:Pitch}
    \end{subfigure}

    \begin{subfigure}{0.75\textwidth}
        \centering
        \includegraphics[width=\textwidth]{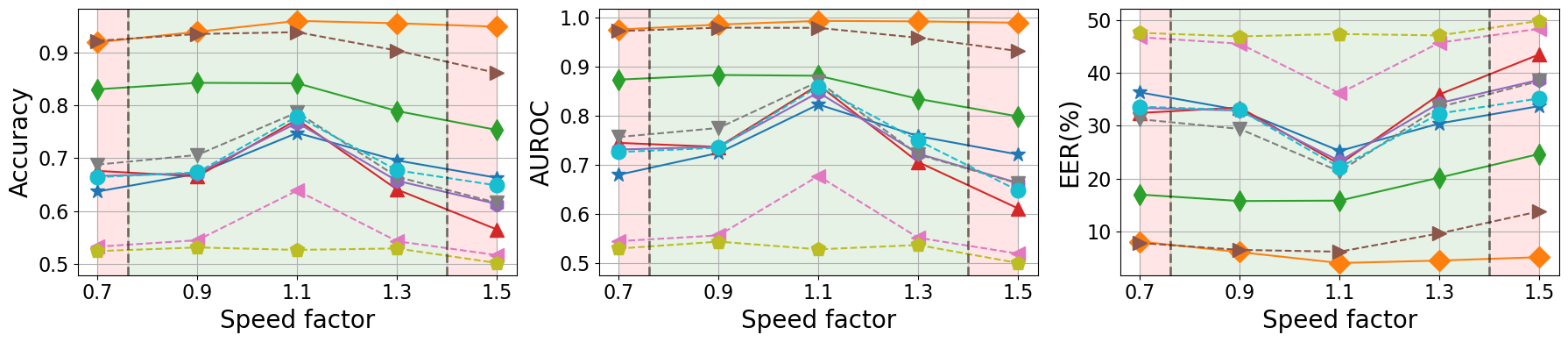}
        \caption{Time stretch}
        \label{fig:time_stretch}
    \end{subfigure}

        \begin{subfigure}{0.75\textwidth}
        \centering
        \includegraphics[width=\textwidth]{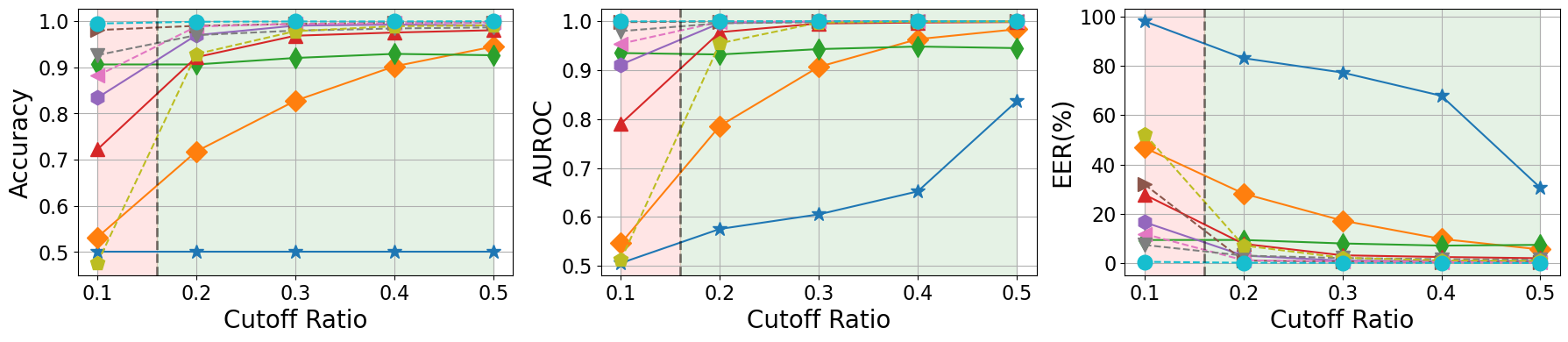}
        \caption{Low-pass filter}
        \label{fig:low_pass}
    \end{subfigure}

    \begin{subfigure}{0.75\textwidth}
        \centering
        \includegraphics[width=\textwidth]{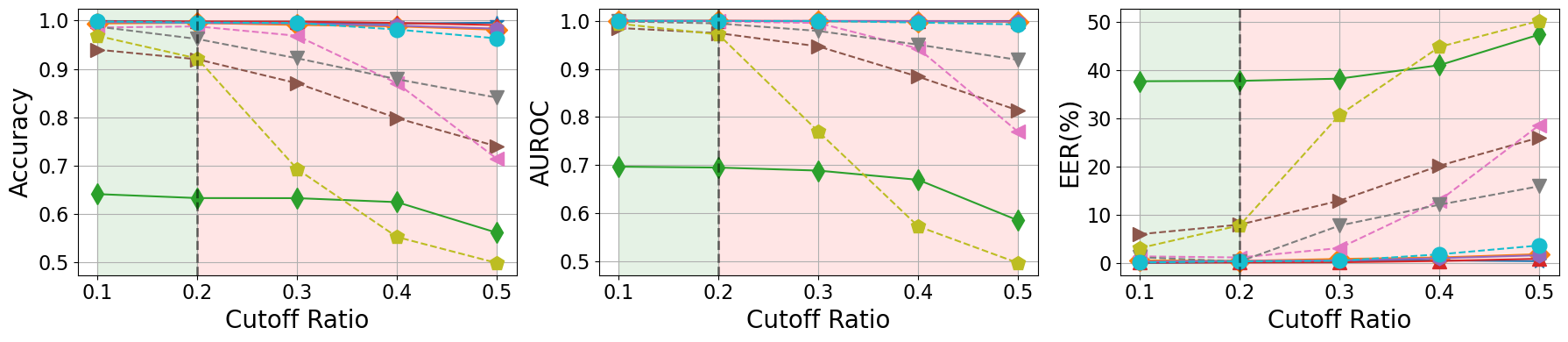}
        \caption{High-pass filter}
        \label{fig:high_pass}
    \end{subfigure}

    \begin{subfigure}{0.75\textwidth}
        \centering
        \includegraphics[width=\textwidth]{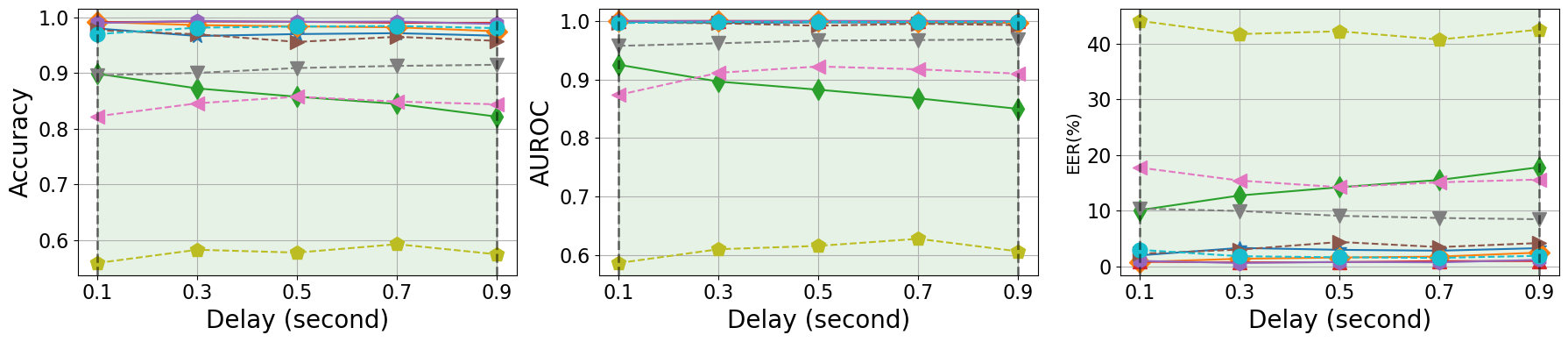}
        \caption{Echo}
        \label{fig:Echo}
    \end{subfigure}

    \begin{subfigure}{0.75\textwidth}
        \centering
        \includegraphics[width=\textwidth]{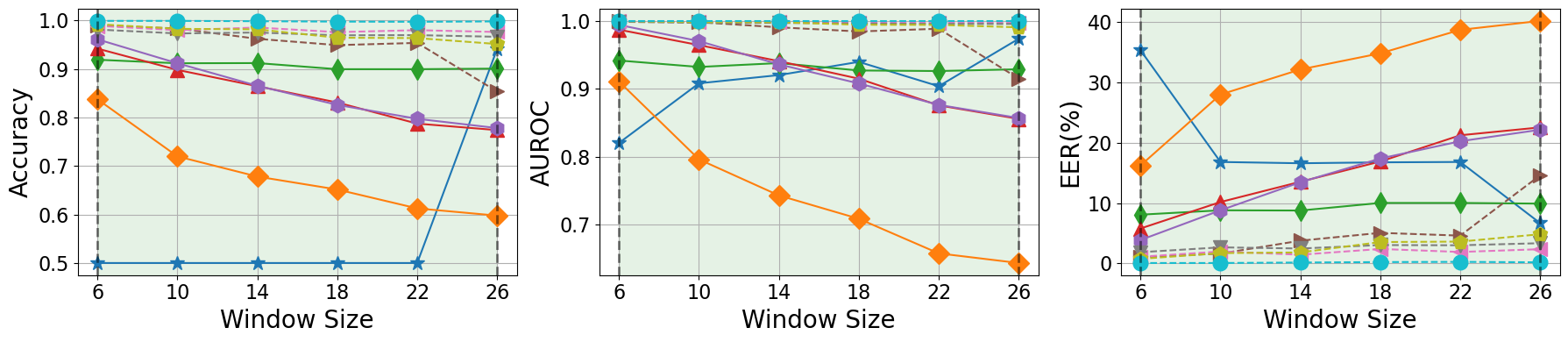}
        \caption{Smooth}
        \label{fig:smooth}
    \end{subfigure}

    \vspace{-0.3cm}
\end{figure*}

\begin{figure*}
    \ContinuedFloat

    \begin{subfigure}{0.75\textwidth}
        \centering
        \includegraphics[width=\textwidth]{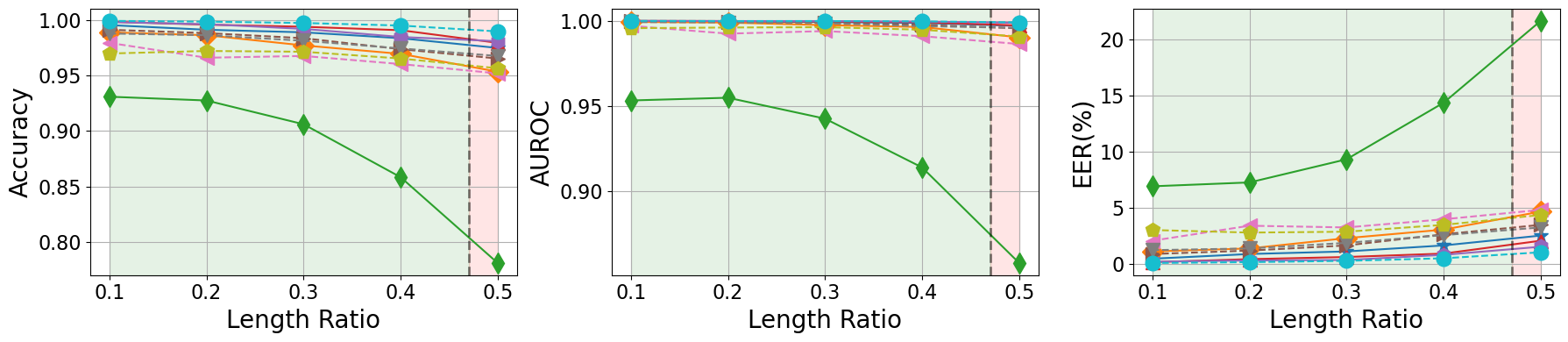}
        \caption{Silence Insertion}
        \label{fig:silence}
    \end{subfigure}
    \begin{subfigure}{0.75\textwidth}
        \centering
        \includegraphics[width=\textwidth]{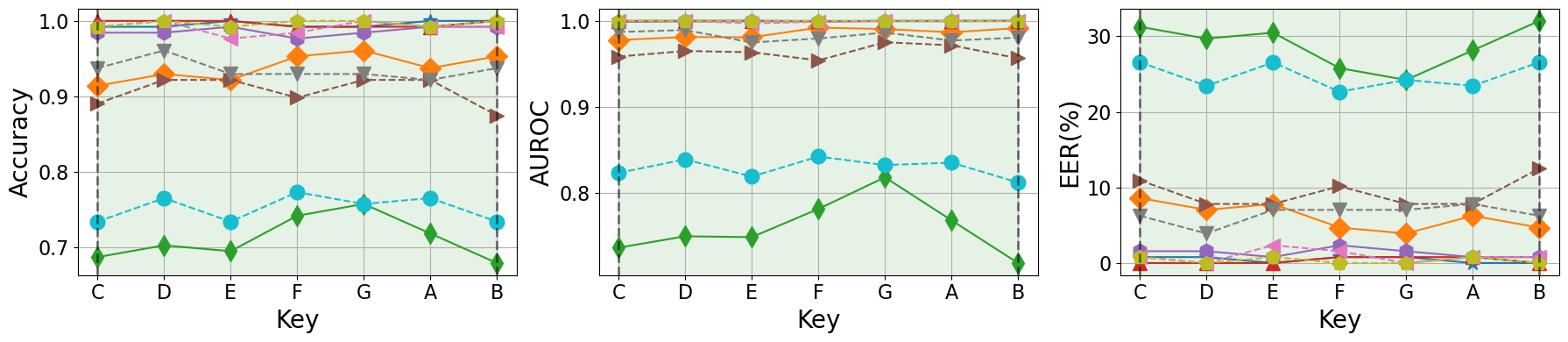}
        \caption{Autotune}
        \label{fig:autotune}
    \end{subfigure}

        \caption{Robustness against different types of modification. Detection performance generally declines as modifications alter critical spectral or temporal features, even when the perceptual audio quality remains high. Foundation models, such as Wave2Vec2BERT and HuBERT, demonstrate stronger robustness across most modification types, while traditional models are more vulnerable to changes, particularly those affecting time and frequency characteristics.}

    \label{fig:modification}
    
\end{figure*}

\subsection{Robustness against Noise Perturbation}
\label{sec:noise}

We begin by evaluating how well audio deepfake detection models perform under noise perturbations, focusing on three common types: Gaussian noise, background noise, and background music. 

\textbf{Figure~\ref{fig:noise_perturbation}} presents the performance of 10 audio deepfake detection models when under Gaussian noise, background noise, and background music at varying SNR levels ranging from 5dB to 40dB. Noise is directly added to the audio waveform to simulate realistic conditions. 

Green-shaded regions in the figure indicate SNR levels where the perturbed audio maintains acceptable quality, defined as having a ViSQOL score of 3 or higher.

As expected, detection performance improves with increasing SNR: models achieve higher accuracy and AUROC and lower EER as the noise level decreases. Higher SNR values correspond to better audio quality with reduced noise interference, facilitating easier detection. At very low SNR levels (e.g., SNR = 5 dB), detection performance drops substantially, highlighting the difficulty in detecting deepfakes when audio is heavily corrupted by noise.

Although detection improves as noise levels decrease, real-world audio often contains non-negligible noise. 
Among the models evaluated, Wave2Vec2BERT consistently demonstrates strong performance, even under severe noise conditions such as 5 dB SNR. It achieves lower EER and higher AUROC than other models across all noise types. \textbf{More broadly, foundation models, including Wave2Vec2BERT, Whisper, and HuBERT,
tend to outperform traditional models like LFCC-LCNN and ResNet\_Spec., especially when the noise is intense}. This robustness likely stems from the extensive pretraining of foundation models on large and diverse audio datasets, which enables them to extract more robust and noise-resilient features.

In contrast to Gaussian and background noise, adding background music to the audio even at a low SNR of 5 dB, the ViSQOL score remains above 3, indicating high perceptual quality. Most models demonstrate strong robustness across all SNR levels, with the exception of RawNet2, CLAP, and Wave2Vec2, which exhibit noticeable performance degradation at lower SNRs. These findings highlight that model robustness varies significantly across different types of noise. 

\subsection{Robustness against Modification}
\label{sec:modification}

We then evaluate how audio deepfake detection models perform under various audio modifications. The results are shown in
\textbf{Figure~\ref{fig:modification}}. Figure~\ref{fig:Pitch} and Figure~\ref{fig:time_stretch} demonstrate the effects of Pitch shifting and Time stretching. We observe that these two types of modifications pose significant challenges for most audio deepfake detection models. Pitch shifts distort subtle spectral features critical for detecting deepfakes, while time stretching disrupts temporal patterns that many models rely on for accurate detection. Even when these modifications do not substantially degrade audio quality,  
they still introduce severe deterioration to audio deepfake detection, even for foundation models that generally present strong robustness.  

Figure~\ref{fig:low_pass} and Figure~\ref{fig:high_pass} present the results for low-pass and high-pass filtering. We observe that detection performance declines as key frequency components are removed, but the impact differs between the two filters. High-pass filtering causes more severe degradation, particularly at higher cutoff ratios, suggesting that many models depend on high-frequency features for detection. For instance, LFCC-LCNN and ResNet\_Spec. show greater resilience to high-pass filtering but perform poorly under low-pass filtering. In contrast, foundation models such as Wave2Vec2, HuBERT, and Whisper are more robust to low-pass filtering but suffer substantial performance drops with aggressive high-pass filtering. Notably, models like Wave2Vec2BERT, AASIST, and RawGATST exhibit strong robustness under both filtering types, likely due to their ability to capture information across a broad frequency range.

Figure~\ref{fig:Echo} illustrates the impact of echo perturbations, with delays ranging from 0.1 to 0.9 seconds, on audio deepfake detection models. Echo introduces reverberations that alter the temporal structure of the audio without significantly affecting spectral features. This allows most models to retain high performance, although temporal distortions can affect models sensitive to time-domain alterations, such as Wave2Vec2. The ViSQOL scores remain consistently above 4 across all delay values, ensuring that echo, even at larger delays, does not heavily distort the perceptual quality of the audio. Despite preserved perceptual quality, detection performance declines noticeably, even with a minimal delay of 0.1 seconds. Wave2Vec2 is particularly sensitive, with Accuracy dropping to 0.558. HuBERT experiences a moderate decline (Accuracy = 0.823), while Whisper shows stronger resilience (Accuracy = 0.9). As delay increases, models like RawNet2 exhibit a gradual decline, reflecting vulnerability to longer delays. In general, models relying on frequency-domain features, such as Wave2Vec2BERT, LFCC-LCNN, and ResNet\_Spec, show better robustness against echo modification. Hybrid models like AASIST and RawGATST, which leverage both temporal and spectral features, also exhibit strong resilience.

By comparison, smoothing (Figure~\ref{fig:smooth}) progressively blurs fine-grained details in the waveform and reduces high-frequency components in the signal as the window size increases. Foundation models with strong feature extraction capabilities, such as Wave2Vec2BERT and HuBERT, are less affected by smoothing. In contrast, traditional models like AASIST and RawGATST experience performance degradation as smoothing impacts both the temporal and spectral domains. 

Silence insertion (Figure~\ref{fig:silence}) evaluates robustness by masking short audio segments as silence at varying ratios. Unlike pitch or spectral modifications, silence insertion preserves local spectral characteristics but breaks global temporal continuity. Detection performance degrades consistently as the insertion ratio increases. Detectors such as Wave2Vec2BERT, AASIST, and RawGATST exhibit stronger robustness under mild silence insertion, likely because their input representations emphasize spectral features and are less exclusively dependent on temporal features. In contrast, RawNet2 shows more significant performance degradation, indicating its higher sensitivity to temporal fragmentation. Nevertheless, even the more robust models suffer notable performance drops at higher insertion ratios, as excessive temporal disruption exceeds their learned invariances.

Autotune (Figure~\ref{fig:autotune}) alters pitch by quantizing it to musical keys while largely preserving phonetic content and perceptual quality. Its impact is not uniform across models. Detectors such as LFCC-LCNN, AASIST, RawGATST, HuBERT, and Whisper remain relatively stable across keys. In contrast, RawNet2 and Wave2Vec2BERT consistently exhibit lower performance and higher EER, suggesting a stronger dependence on fine-grained pitch dynamics or raw waveform characteristics that are disrupted by pitch quantization.

 Overall, \textbf{our evaluation shows that no model consistently maintains robustness across all modification types}. Models' detection performances vary depending on their reliance on spectral or temporal features. In particular, modifications like pitch shifting, time stretching, and echo, despite their minimal impact on perceptual audio quality, can severely impair detection accuracy. These findings highlight the critical need to address these types of subtle modifications during model development, as their nuanced effects can undermine detection performance even when the audio quality remains high.

\subsection{Robustness against Compression}
\label{sec:compression}

\begin{figure*}[t]
    \centering

    \begin{subfigure}{0.75\textwidth}
        \centering
        \includegraphics[width=\textwidth]{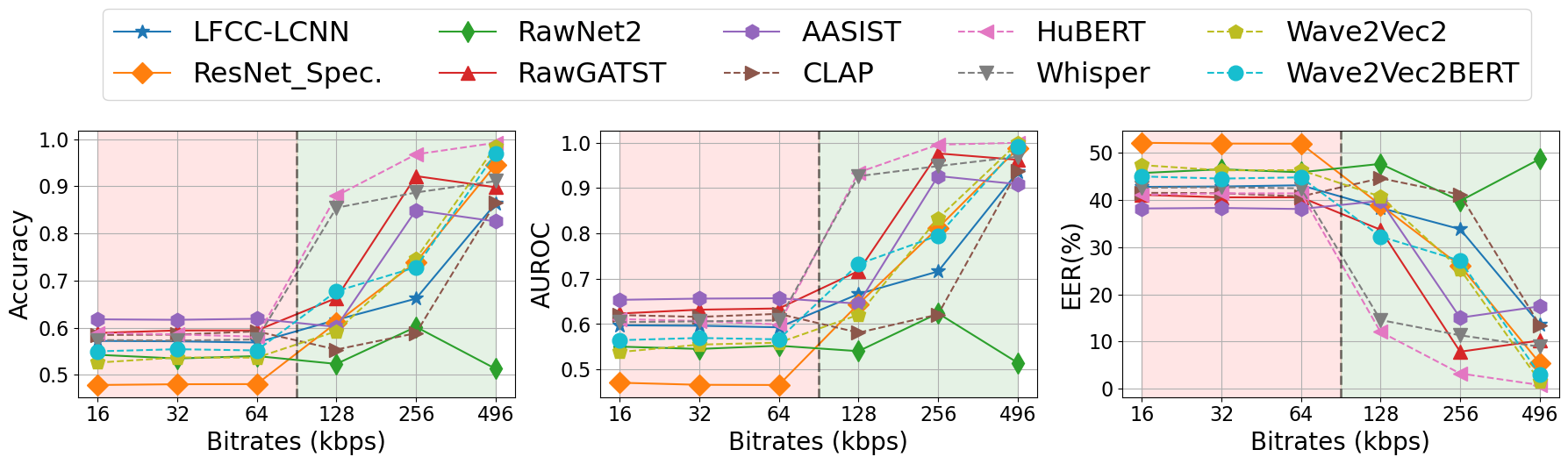}
        \caption{Opus}
        \label{fig:opus}
    \end{subfigure}

    \begin{subfigure}{0.75\textwidth}
        \centering
        \includegraphics[width=\textwidth]{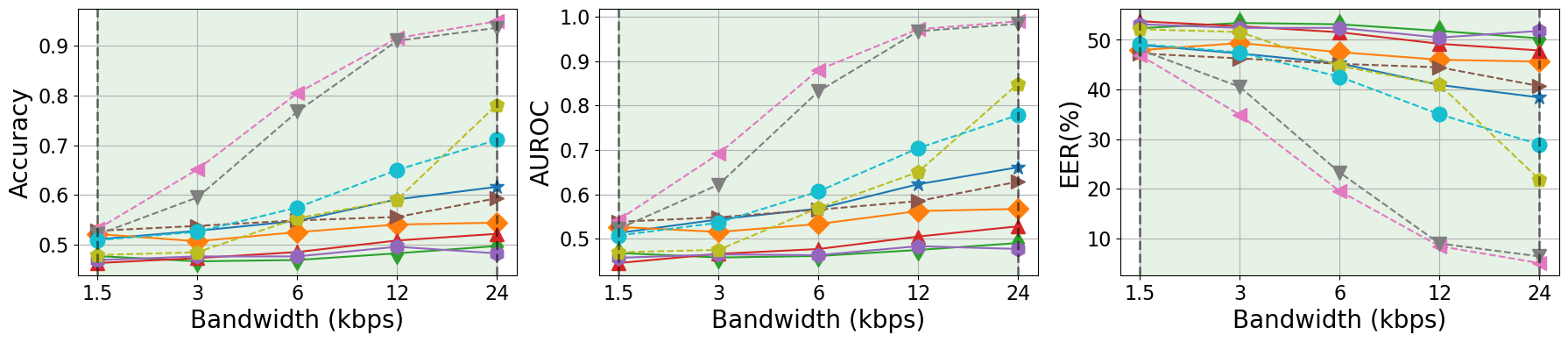}
        \caption{Encodec}
        \label{fig:encodec}
    \end{subfigure}

    \begin{subfigure}{\textwidth}
        \centering
        \includegraphics[width=0.75\textwidth]{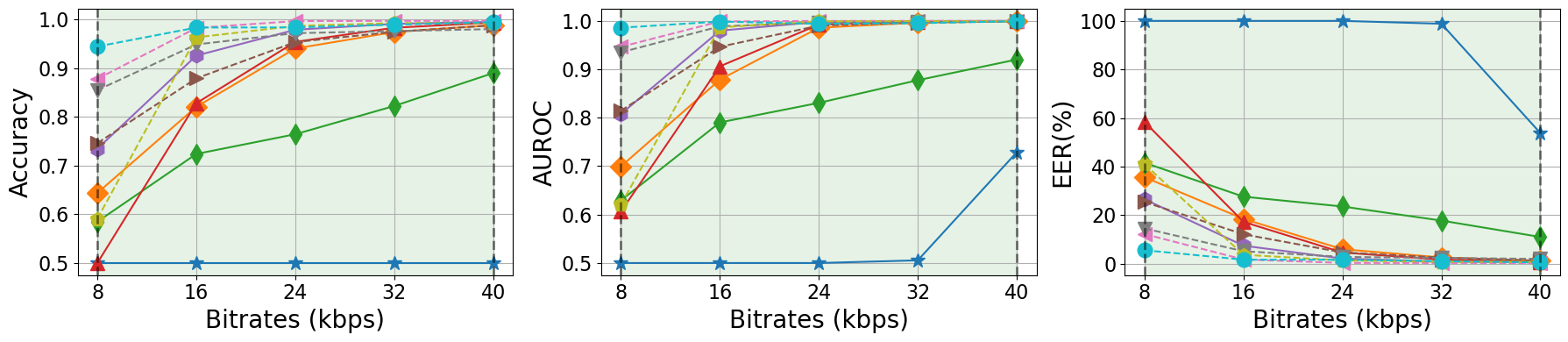}
        \caption{MP3}
        \label{fig:mp3}
    \end{subfigure}

    \begin{subfigure}{\textwidth}
        \centering
        \includegraphics[width=0.75\textwidth]{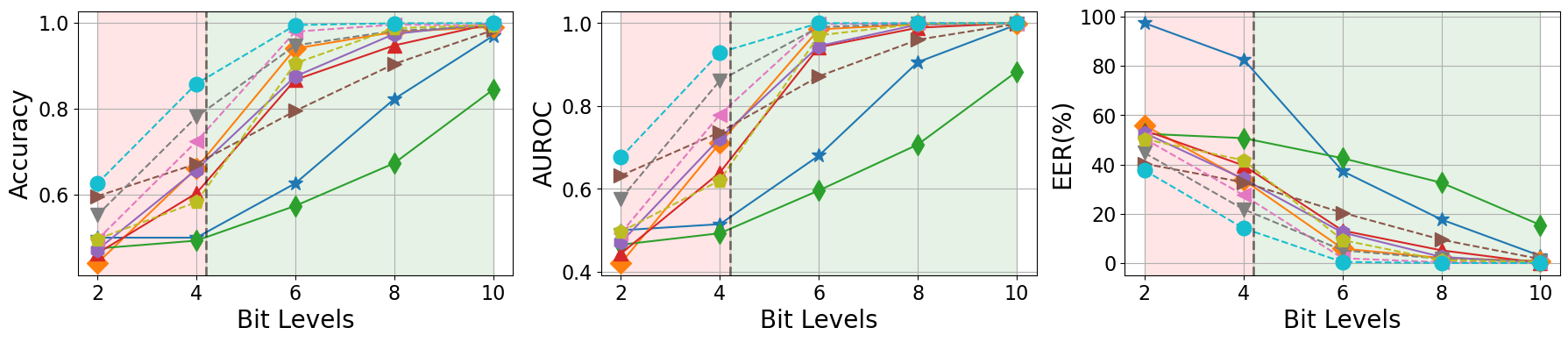}
        \caption{Quantization}
        \label{fig:quantization}
    \end{subfigure}

    \begin{subfigure}{\textwidth}
        \centering
        \includegraphics[width=0.75\textwidth]{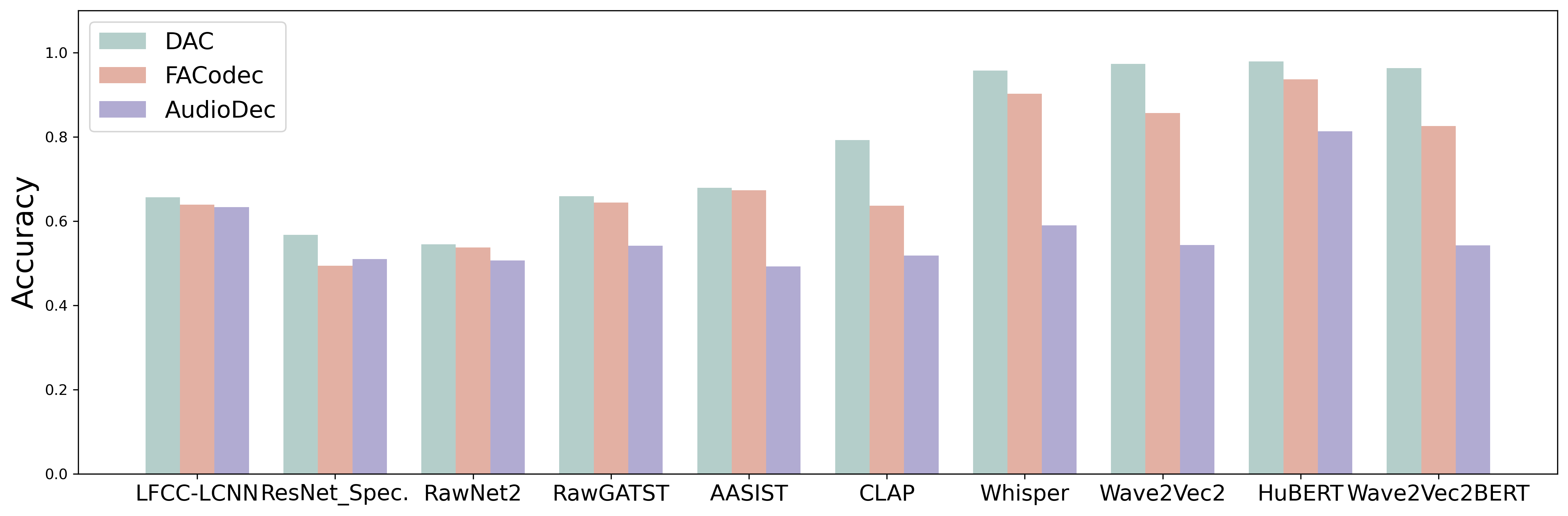}
        \caption{Accuracy on other types of neural codecs.}
        \label{fig:other_codec}
    \end{subfigure}

    \caption{Robustness against Compression. Although higher audio quality generally leads to better detection performance, compression can introduce subtle artifacts that undermine model robustness. In particular, neural codecs like Encodec, which maintain high perceptual audio quality even at low bandwidth, still cause substantial detection performance drop.
    }
\label{fig:compression}
\vspace{-0.4cm}
\end{figure*}

Next, we evaluate the robustness of audio deepfake detection models under various types of audio compression methods.

\textbf{Figure~\ref{fig:compression}} presents the audio deepfake detection results for various compression corruptions. Our results show that \textbf{despite maintaining high perceptual audio quality, compression significantly impairs detection performance.}
As shown in Figure~\ref{fig:encodec} and Figure~\ref{fig:mp3}, both Encodec and MP3 preserve good audio quality across different compression levels. However, detection models experience considerable performance degradation, especially with Encodec. As shown in Figure~\ref{fig:mp3}, even when audio is compressed to as low as 8 kbps—while still maintaining high perceptual quality—most models exhibit a substantial drop in detection accuracy.

Figure~\ref{fig:opus}, Figure~\ref{fig:quantization}, and Figure~\ref{fig:other_codec} present the robustness of audio deepfake detectors against Opus compression, quantization, and three types of neural codecs (DAC~\cite{kumar2024high}, FACodec~\cite{ju2024naturalspeech}, and AudioDec~\cite{wu2023audiodec}), respectively. Unlike Encodec, DAC, FACodec, and AudioDec support only a single fixed compression rate. For these codecs, we use their default configurations to compress and decompress the audio. For Opus, even at a high bitrate of 256 kbps, detection performance remains limited. A similar trend is observed in Figure~\ref{fig:quantization}, where reducing audio resolution to 2, 4, or 6 bits results in significant performance degradation. Notably, models are especially vulnerable to neural codecs, as illustrated in Figure~\ref{fig:other_codec}, suggesting that such advanced compression techniques can severely disrupt the features relied on for accurate deepfake detection.

Overall, the disconnection between high perceptual quality and model robustness of audio deepfake detection suggests that existing detection models rely on features that are easily disrupted by compression. These subtle artifacts, which are imperceptible to human listeners, are often essential for accurate detection and are lost during compression, leading to degraded model performance. 
Although foundation models exhibit better robustness, their performance still drops noticeably under extreme compression conditions, such as low MP3 bitrates or narrow Encodec bandwidths. Given the widespread use of audio compression in practical settings, this vulnerability poses a serious challenge for deploying deepfake detection models in real-world environments.

Across the three categories of perturbations evaluated, we observe that while most models exhibit strong robustness to noise perturbations, they are significantly more susceptible to modifications and compression. Detection performance degrades significantly when perturbations interfere with the specific feature domains, such as spectral or temporal features, that models rely on. In particular, compression artifacts introduced by neural codecs are especially detrimental. As neural codecs are increasingly adopted in real-world applications, this vulnerability raises critical concerns for the practical reliability of current audio deepfake detection systems.

\begin{figure}[!t]
    \centering
    \includegraphics[width=0.9\columnwidth]{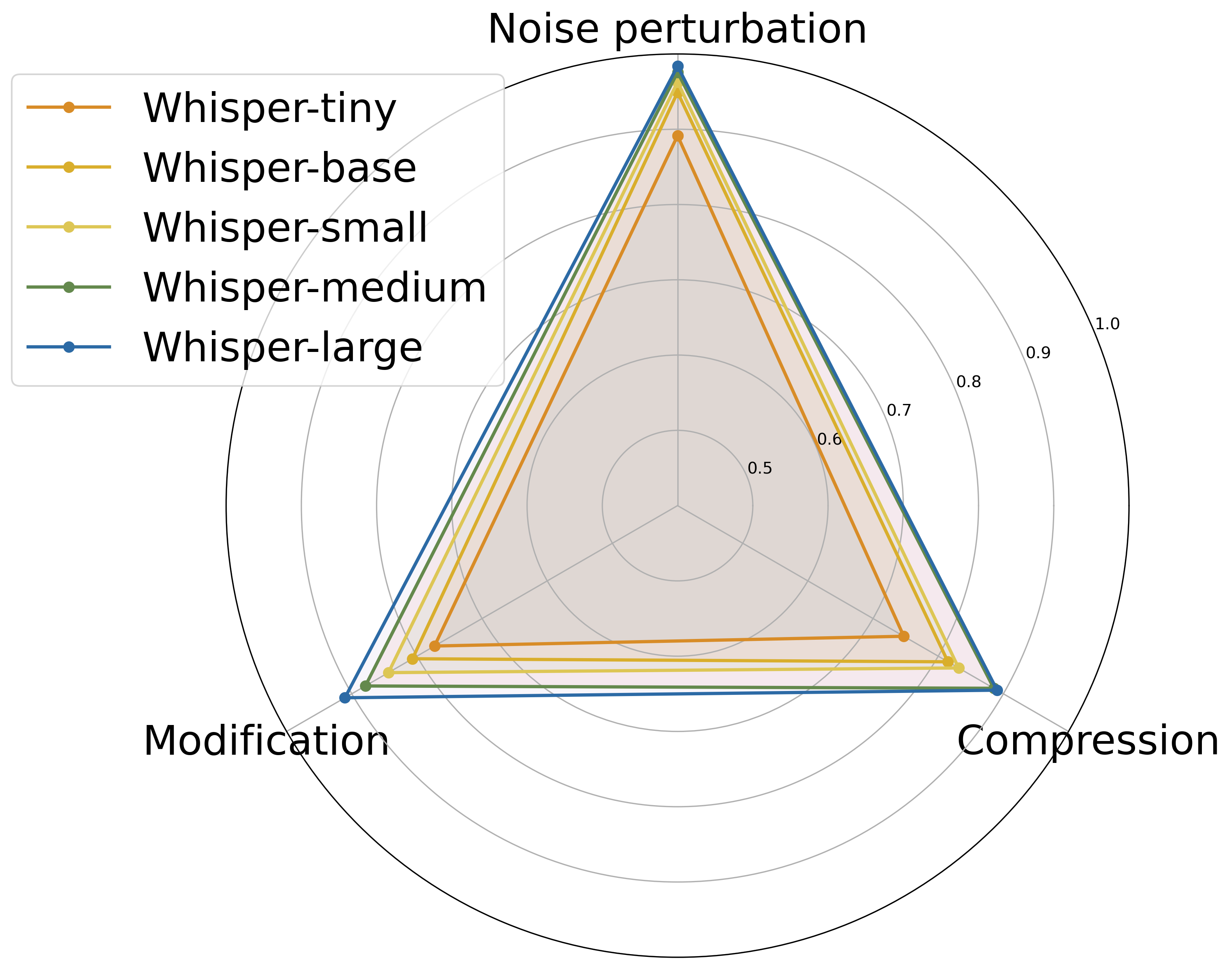}
    \caption{Robustness of different sizes of Whisper models under noise perturbations, modifications, and compression.}
    \label{fig:whisper}
\end{figure}

\subsection{Robustness v.s. Model Size in Foundation Models}
\label{sec:model_size}

Existing work~\cite{li2024sonar} has shown that increasing model size can enhance the generalization ability of audio deepfake detectors across diverse fake audio distributions. To further explore the impact of model size on the robustness, we examine the Whisper~\cite{radford2023robust} model family, which is a widely used speech foundation model with multiple size variants(i.e., tiny, base, small, medium, and large), making it well-suited for studying scaling effects. We fine-tune 
each model version under identical settings and evaluate their detection performance under various audio corruptions. 
 the Whisper family.

\textbf{Figure~\ref{fig:whisper}} presents the average detection accuracies for Whisper model of varying sizes under the three categories of audio corruptions. \textbf{As model size increases, detection performance improves consistently, suggesting that larger foundation models learn more robust and generalizable audio features for audio deepfake detection.} For example, Whisper-large and Whisper-medium not only achieve higher accuracy but also maintain more balanced robustness across all three corruption categories. In contrast, smaller models like Whisper-tiny and Whisper-base show lower accuracy and heightened sensitivity to certain corruptions. While larger models deliver stronger robustness, their computational and storage demands must be carefully weighed in practical applications to balance performance and resource constraints.

\subsection{Improving robustness by augmenting selected audio perturbations}
\label{sec:augmentation}

\begin{figure}[t]
    \centering
    \includegraphics[width=\columnwidth]{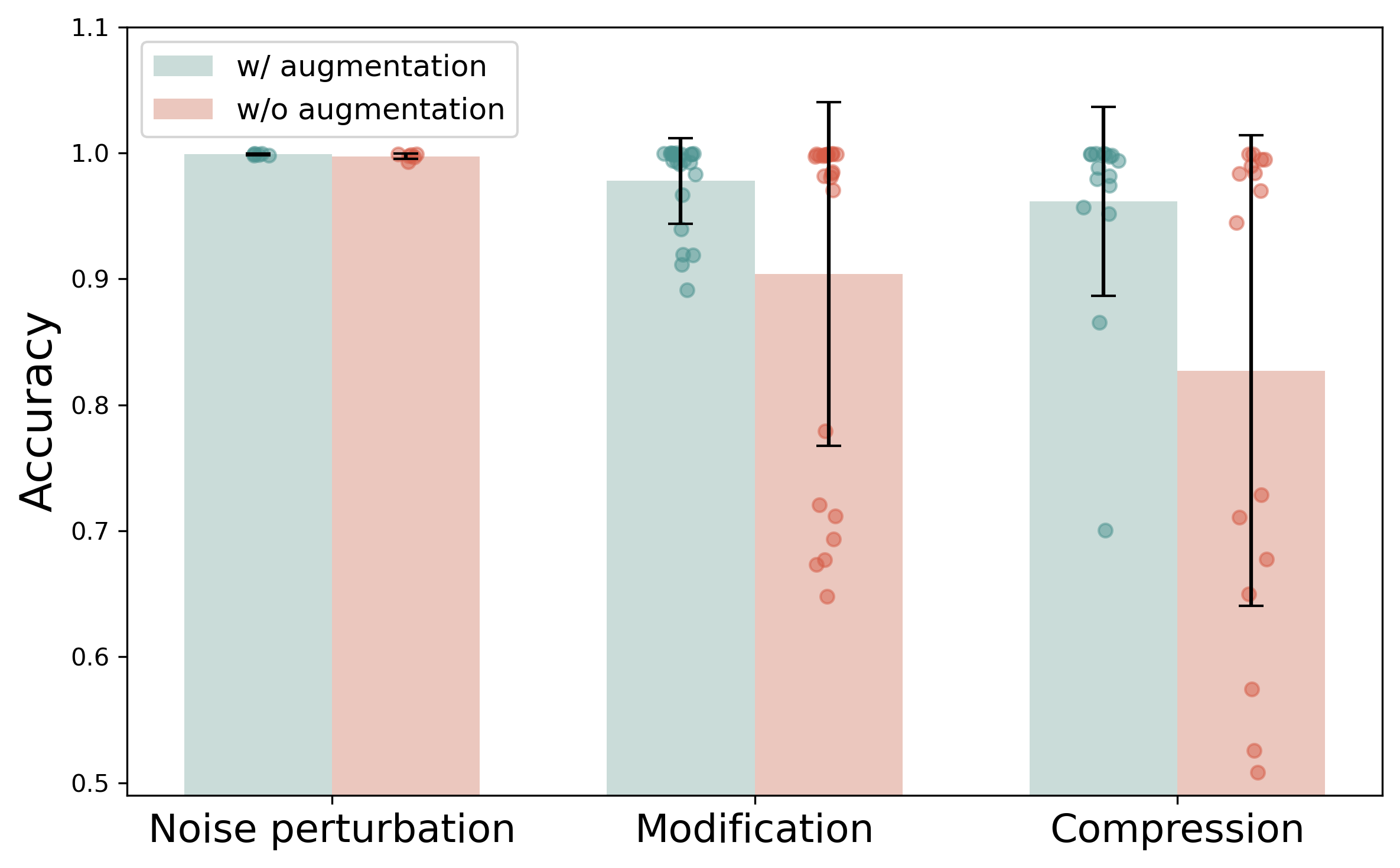}
    \caption{Robustness of Wave2Vec2BERT trained w/ and w/o data augmentation against common corruptions. Although the model without augmentation already shows strong robustness to noise perturbations, incorporating data augmentation further improves detection accuracy and reduces performance variance, highlighting the effectiveness of data augmentation in improving the adaptability of audio deepfake detection models for practical deployment scenarios.}
    \label{fig:augmentation}
\end{figure}

Building on prior work that highlights data augmentation as an effective strategy for improving neural network robustness~\cite{hendrycks2019augmix,rebuffi2021data}, we investigate its impact on enhancing the robustness of audio deepfake detectors against common corruptions. We focus on Wave2Vec2BERT, which is the most robust model identified in our earlier evaluations, and fine-tune it using four particularly challenging perturbations: pitch shifting, time stretch, Opus and Encodec compression. These augmentations are applied during fine-tuning to assess their effect on detection performance under corrupted audio conditions.

\textbf{Figure~\ref{fig:augmentation}} compares the detection accuracy of Wave2Vec2BERT models fine-tuned with and without data augmentation. While the model without augmentation already demonstrates robustness to noise perturbations, data augmentation further reduces performance variance and improves overall robustness. Notable improvements are observed for modification-based corruptions such as pitch shifting and time stretching, and with the most significant gains seen under compression artifacts. These results demonstrate that \textbf{incorporating realistic perturbations during training can significantly improve model generalization and make audio deepfake detection more reliable in real-world applications.}

\begin{figure}[t]
    \centering
    \includegraphics[width=0.9\columnwidth]{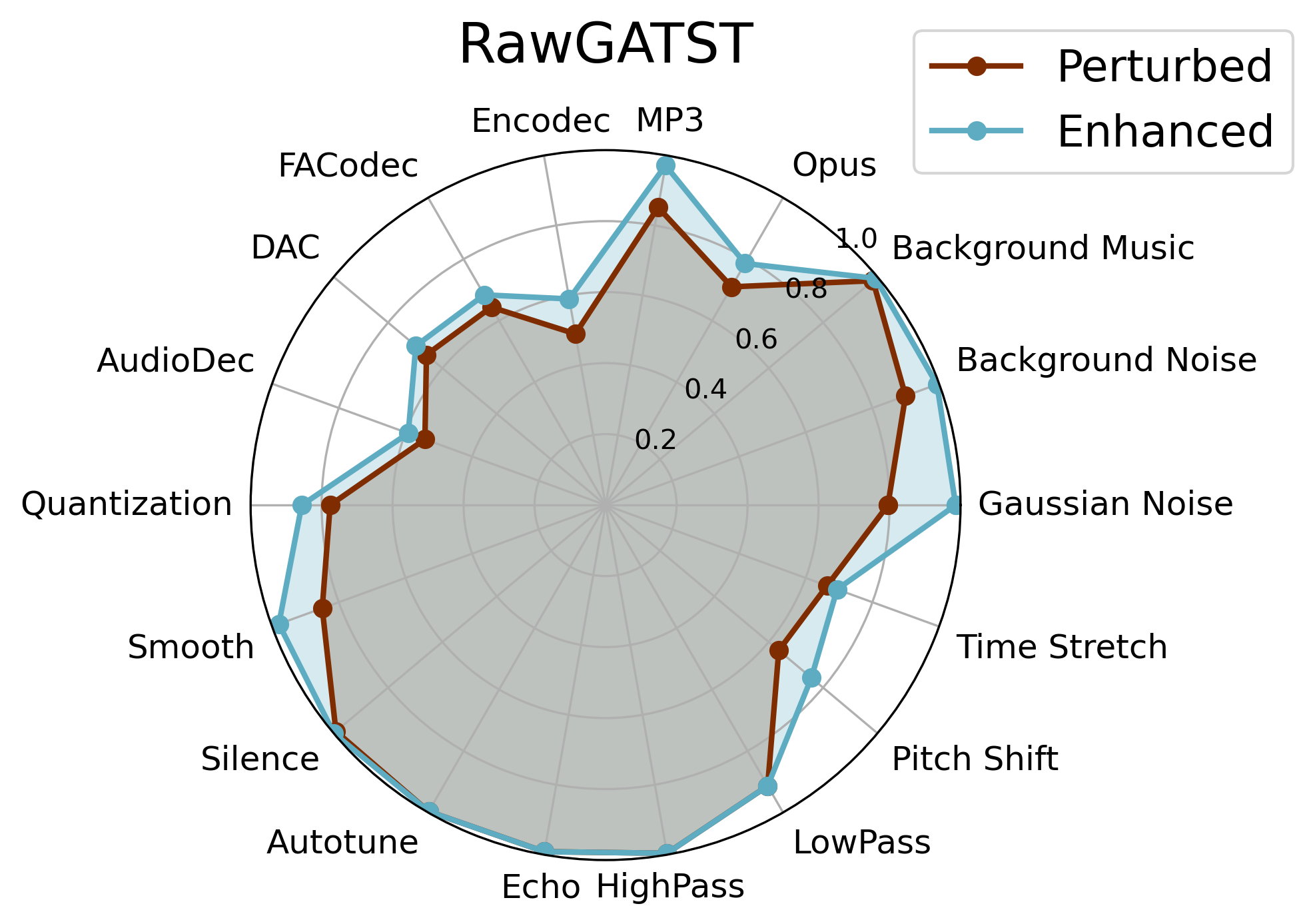}
    \caption{Detection performance for each type of perturbations after applying speech enhancement to corrupted audio before inference.}
    \label{fig:speech_enhance_rawgatst}
\end{figure}

\subsection{Boosting robustness using speech enhancement}
\label{sec:speech_enhancement}

\begin{figure*}[!ht]
    \centering

    \begin{subfigure}{0.28\textwidth}
        \centering
        \includegraphics[width=\textwidth]{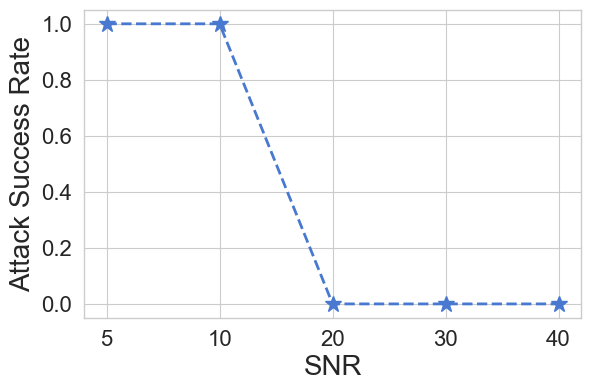}
        \caption{Background Noise}
    \end{subfigure}
    \begin{subfigure}{0.28\textwidth}
        \centering
        \includegraphics[width=\textwidth]{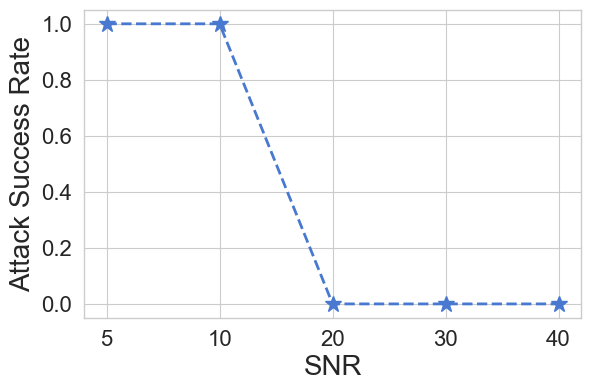}
        \caption{Background Music}
    \end{subfigure}
    \begin{subfigure}{0.28\textwidth}
        \centering
        \includegraphics[width=\textwidth]{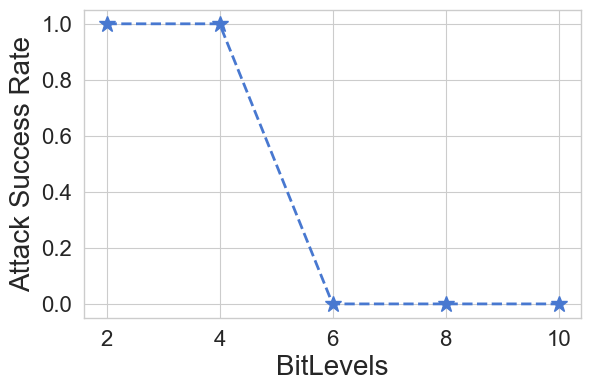}
        \caption{Quantization}
    \end{subfigure}

    \begin{subfigure}{0.28\textwidth}
        \centering
        \includegraphics[width=\textwidth]{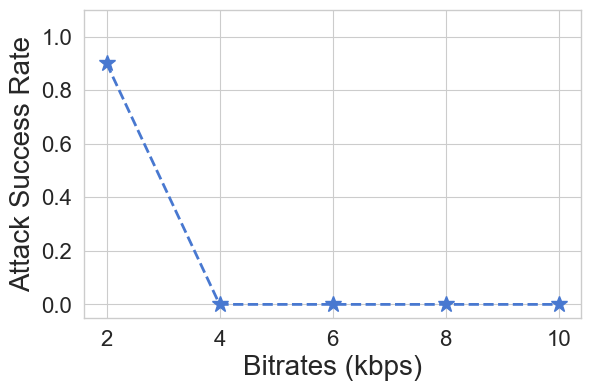}
        \caption{MP3}
    \end{subfigure}
    \begin{subfigure}{0.28\textwidth}
        \centering
        \includegraphics[width=\textwidth]{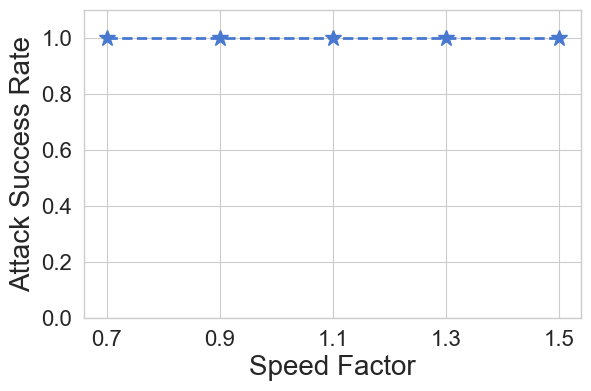}
        \caption{Time Stretch}
    \end{subfigure}
    \begin{subfigure}{0.28\textwidth}
        \centering
        \includegraphics[width=\textwidth]{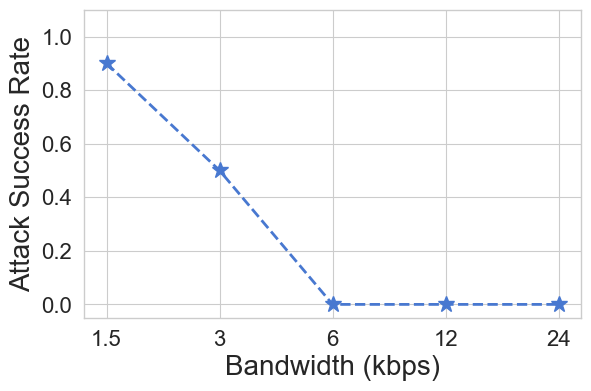}
        \caption{Encodec}
    \end{subfigure}

\caption{Attack success rate on bypassing ElevenLabs' detector.}
\label{fig:elevenlabs}
\end{figure*}

In addition to data augmentation, we investigate whether speech enhancement can improve the robustness of audio deepfake detection. For each corrupted audio sample, we apply a commercial enhancement tool, Adobe Podcast Enhance\footnote{\url{https://podcast.adobe.com/en/enhance}} to improve audio quality before inference. \textbf{Figure~\ref{fig:speech_enhance_rawgatst}} shows the results for RawGATST (The results for other 9 models can be found in \textbf{Figure \ref{fig:speech_enhancement}} in \textbf{Appendix~\ref{appendix:speech_enhance}}), with detection accuracy averaged across all severity levels for each perturbation type. 

Our findings show that speech enhancement significantly improves robustness, particularly under noise-based perturbations and MP3 compression. While gains for other modification and compression types are more modest, consistent improvements are still observed. These results suggest that applying speech enhancement prior to detection can help reduce the negative impact of various audio corruptions, particularly in noisy or low-quality conditions.

\subsection{Bypassing commercial audio watermark detection}

We further use \textit{AudioPerturber} to evaluate how real-world audio corruptions may affect the effectiveness of a commercial audio watermarking system. Specifically, we generate 10 speech samples using ElevenLabs for each perturbation type and each severity level, respectively. We then evaluate these perturbed samples using ElevenLabs’ proprietary detector, which detects whether an audio clip was generated by their system. Specifically, ElevenLabs’ detector outputs one of five confidence levels for each audio input: Very Likely, Likely, Uncertain, Unlikely, and Very Unlikely. We consider the last three categories (i.e., Uncertain, Unlikely, and Very Unlikely) as detection failures, and adopt the attack success rate (ASR) as the evaluation metric. Figure~\ref{fig:elevenlabs} summarizes the perturbation types that bypassed the detector at least once across different severity levels. Note that AudioDec also successfully bypasses the detector, achieving an ASR of 0.6. However, since it offers only one severity level, it is excluded from Figure~\ref{fig:elevenlabs}.

Interestingly, the detector demonstrates strong robustness overall. Out of 18 perturbation types, only background noise, background music, high-pass filtering, MP3, Encodec, AudioDec, and quantization successfully bypassed the detector, and only at higher severity levels, where the audio quality is significantly degraded. However, one notable exception is Time Stretch, which consistently bypassed the detector across all severity levels, suggesting a potential vulnerability that merits further attention.

Overall, the detector exhibits strong robustness. Out of the 18 perturbation types tested, only Background Noise, Background Music, Time Strech, MP3 compression, Encodec, AudioDec, and quantization were able to bypass detection—and primarily at high severity levels where the audio becomes noticeably degraded. However, a notable exception is Time Stretch, which consistently evades detection across all severity levels, indicating a potential vulnerability that warrants further investigation.

\section{Conclusion}
\label{sec:conclusion}
This work systematically evaluates the robustness of 10 SOTA audio deepfake detection models against 18 common real-world corruptions, including noise perturbations, modifications, and compression. We also explored potential strategies to improve detection robustness. Our key findings are as follows: 
\begin{itemize}[leftmargin=*]
    \item While most models exhibit strong robustness to noise perturbations, they struggle significantly with modification and compression. Detection performance is
    particularly vulnerable when corruptions alter the specific feature domains the models rely on, e.g., spectral or temporal features.
    Compression artifacts introduced by neural codecs are particularly harmful. Given the growing adoption of neural codecs in real-world applications, this vulnerability is particularly concerning. 
    \item Foundation models such as Wave2Vec2BERT, HuBERT, and Whisper exhibit greater robustness to various corruptions,  despite not being explicitly trained on them.
    However, their large size and high computational cost limit their feasibility for real-time or resource-constrained deployments. This highlights the need for more efficient and practical solutions that can balance robustness with practicality. 
    \item Increasing model size generally correlates with improved robustness against corruptions. However, the robustness gains diminish beyond a certain scale, suggesting diminishing returns with continued size increases. 
    \item Incorporating corruption-specific data augmentation during training significantly improves model resilience to unseen distortions. Additionally, applying speech enhancement techniques before inference can help mitigate the impact of corruptions, especially under noisy or compressed conditions.
\end{itemize}

To build deepfake detectors that are both accurate and robust, future research should simulate diverse real-world conditions and evaluate models under a broad spectrum of data corruptions, including emerging challenges like neural codec compression.

Beyond audio deepfakes, the insights from our evaluation can extend to other deepfake-affected domains, such as images and videos, which face their own forms of corruption. A consistent focus on robustness, domain-specific corruptions, and generalizable training strategies will be essential for developing reliable deepfake detection systems across media types.

\section*{Acknowledgments}
The authors thank the partial support from the Fordham–IBM Research Fellowship, the Fordham Faculty Research Grant, and the New York Public Interest Technology Regional Network Seed Grant.

\bibliographystyle{ACM-Reference-Format}
\bibliography{main.BibLaTeX}

\appendix

\section{Audio quality}
\label{appendix:audio_quality}

\begin{figure*}[htbp]
    \centering
    \begin{subfigure}[b]{0.28\textwidth}
        \centering
        \includegraphics[width=\textwidth]{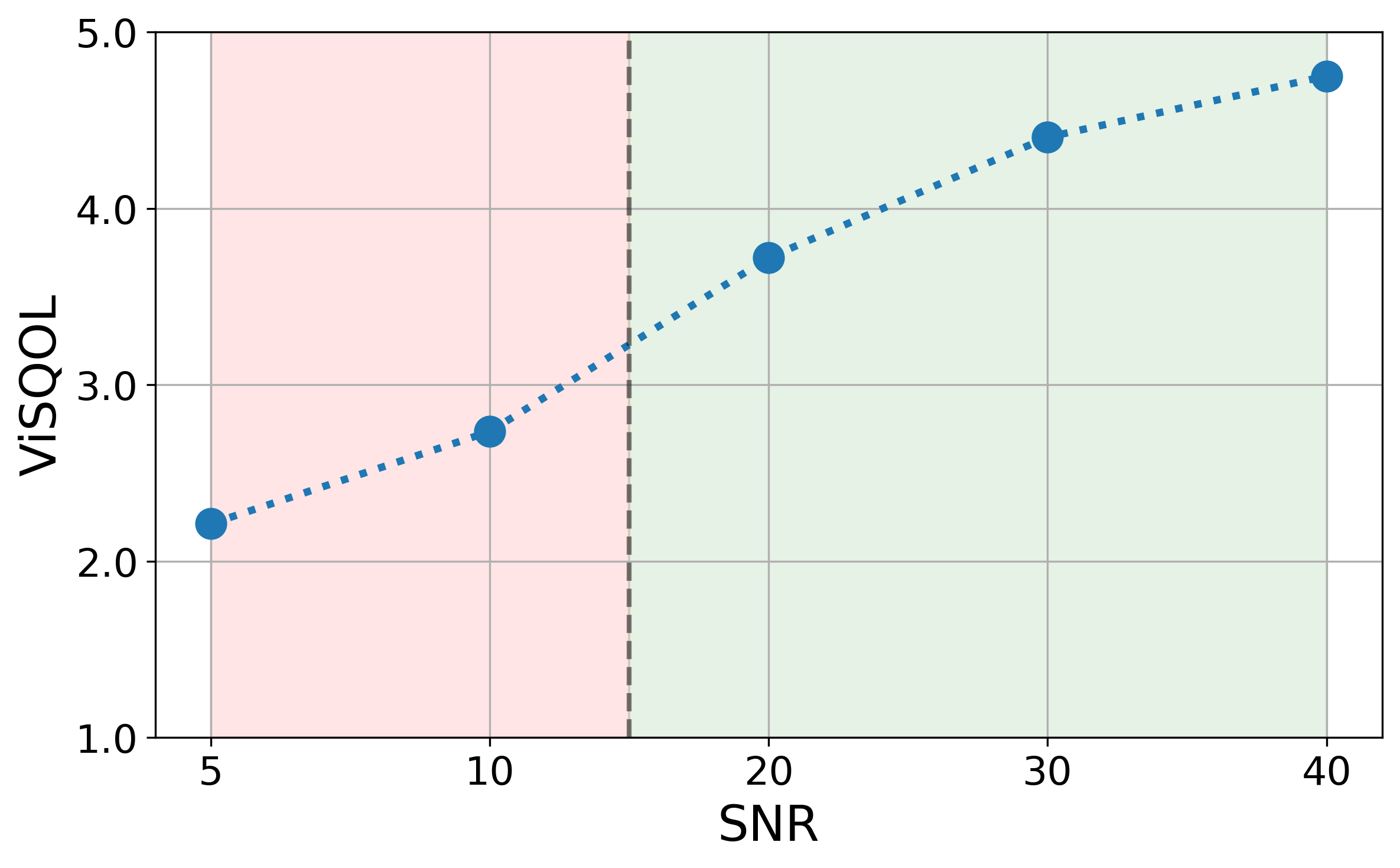}
        \caption{Gaussian noise}
        \label{fig:visqol_gaussian}
    \end{subfigure}
    \begin{subfigure}[b]{0.28\textwidth}
        \centering
        \includegraphics[width=\textwidth]{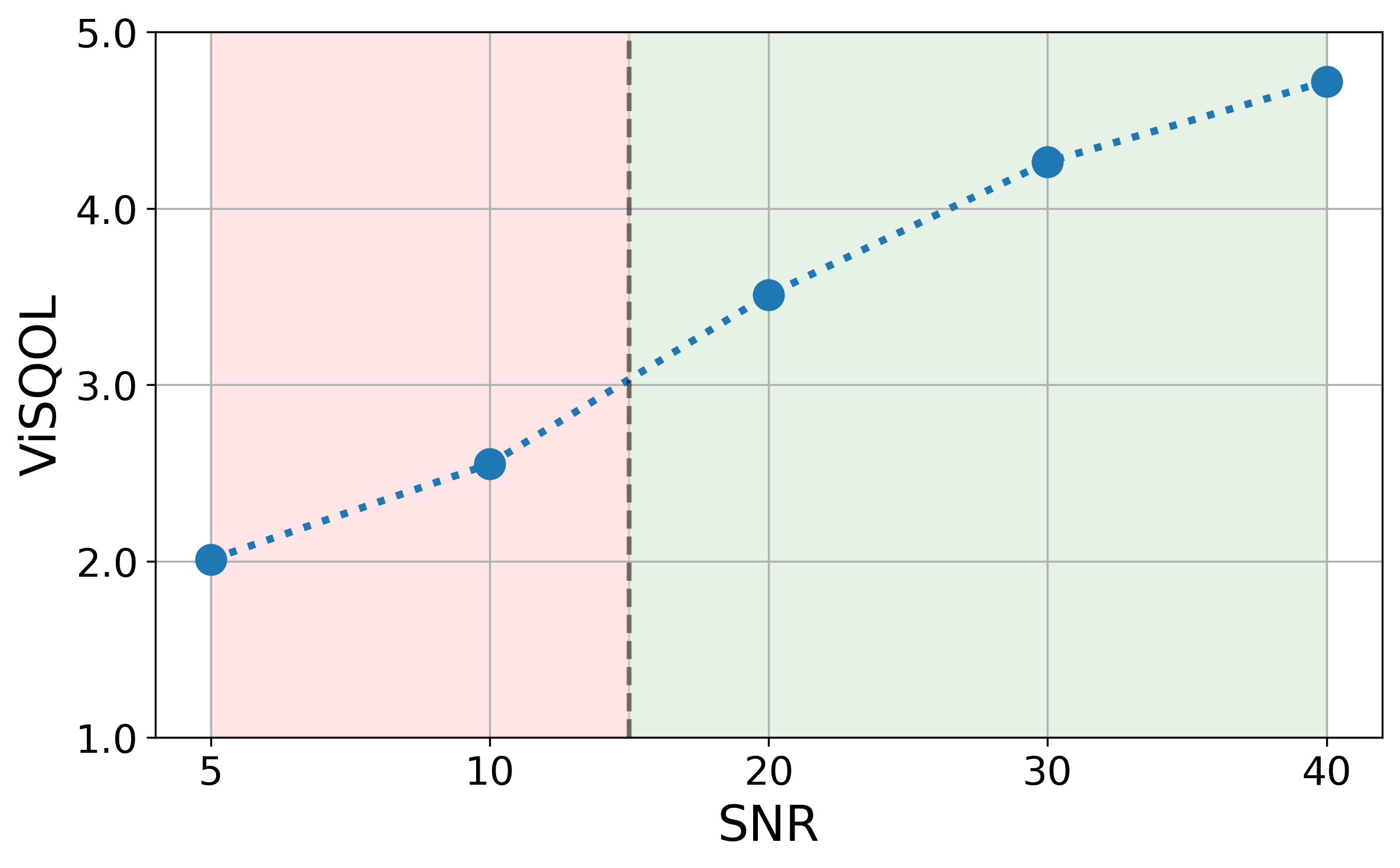}
        \caption{Background noise}
        \label{fig:visqol_bg}
    \end{subfigure}
    \begin{subfigure}[b]{0.28\textwidth}
        \centering
        \includegraphics[width=\textwidth]{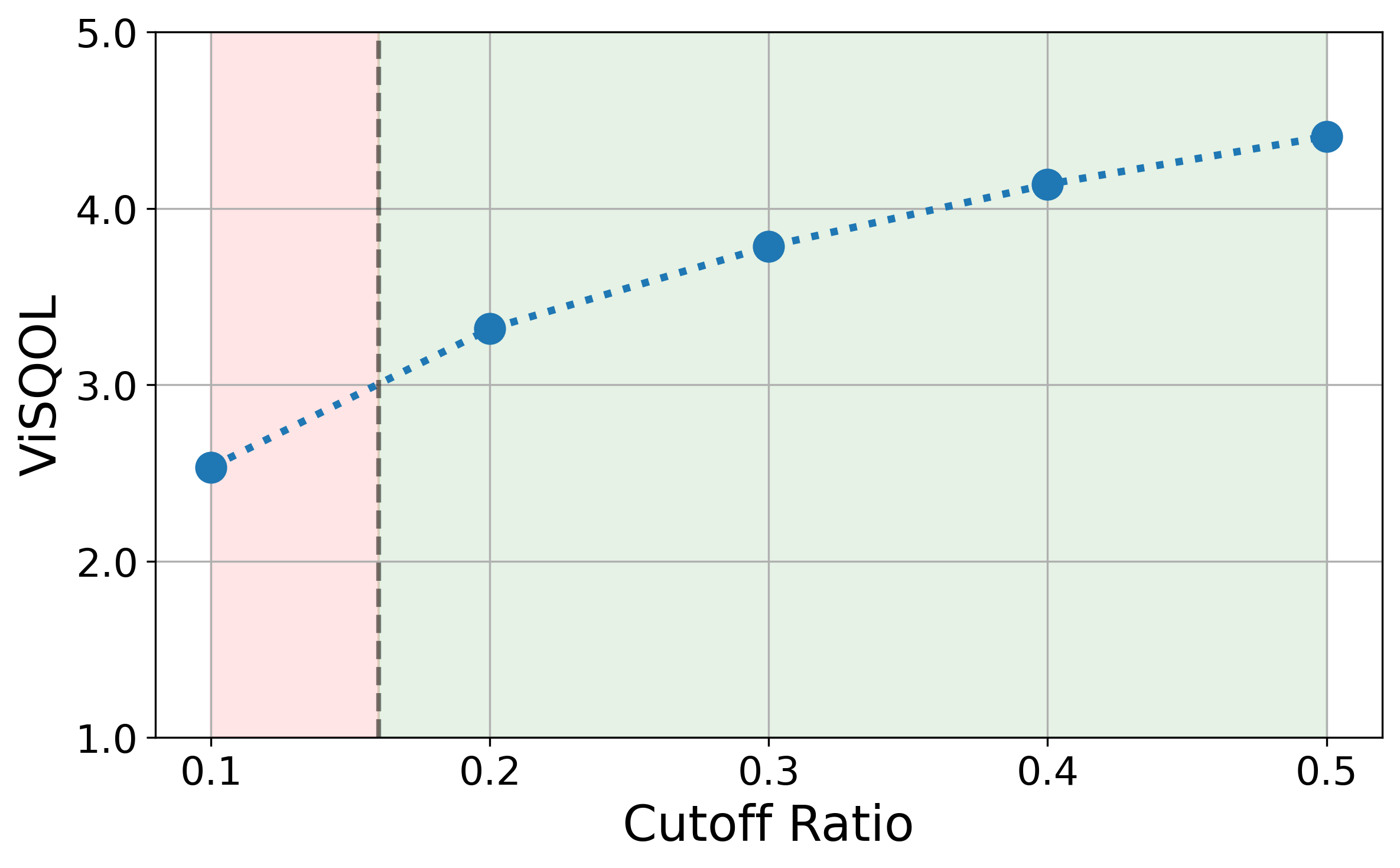}
        \caption{Low-pass filter}
        \label{fig:visqol_low_pass}
    \end{subfigure}
    
    \begin{subfigure}[b]{0.28\textwidth}
        \centering
        \includegraphics[width=\textwidth]{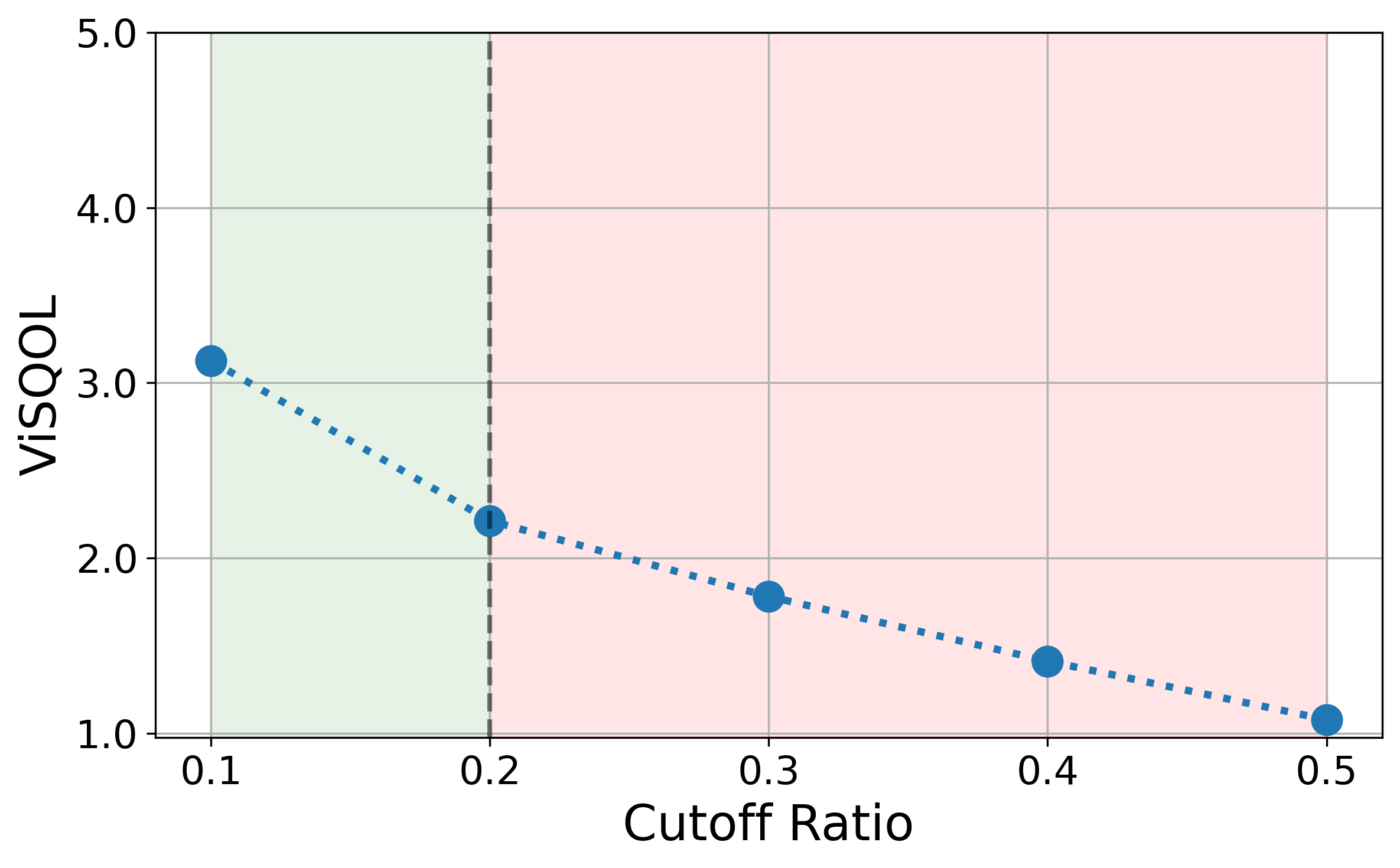}
        \caption{High-pass filter}
        \label{fig:visqol_high_pass}
    \end{subfigure}
    \begin{subfigure}[b]{0.28\textwidth}
        \centering
        \includegraphics[width=\textwidth]{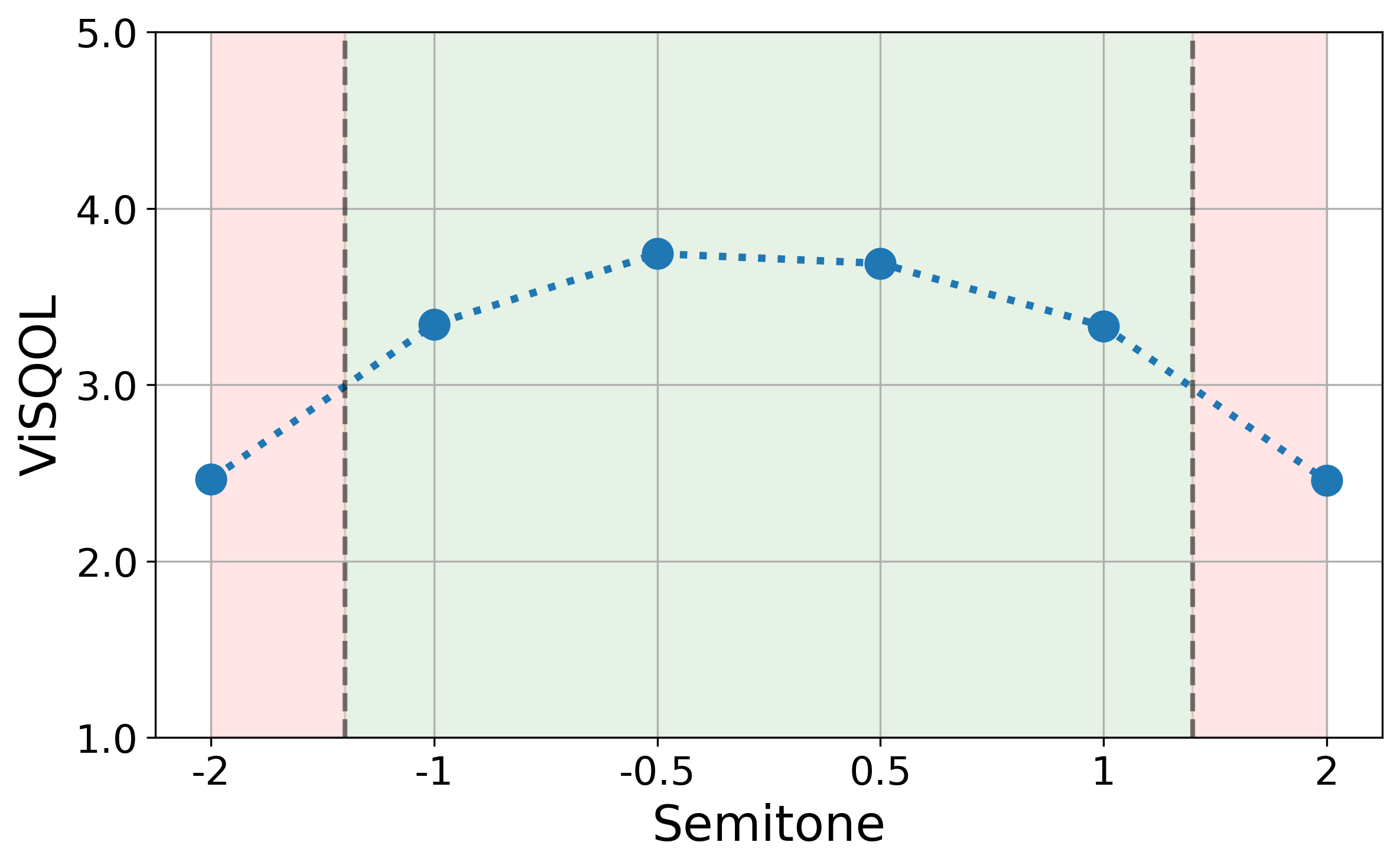}
        \caption{Pitch shifting}
        \label{fig:visqol_pitch}
    \end{subfigure}
    \begin{subfigure}[b]{0.28\textwidth}
        \centering
        \includegraphics[width=\textwidth]{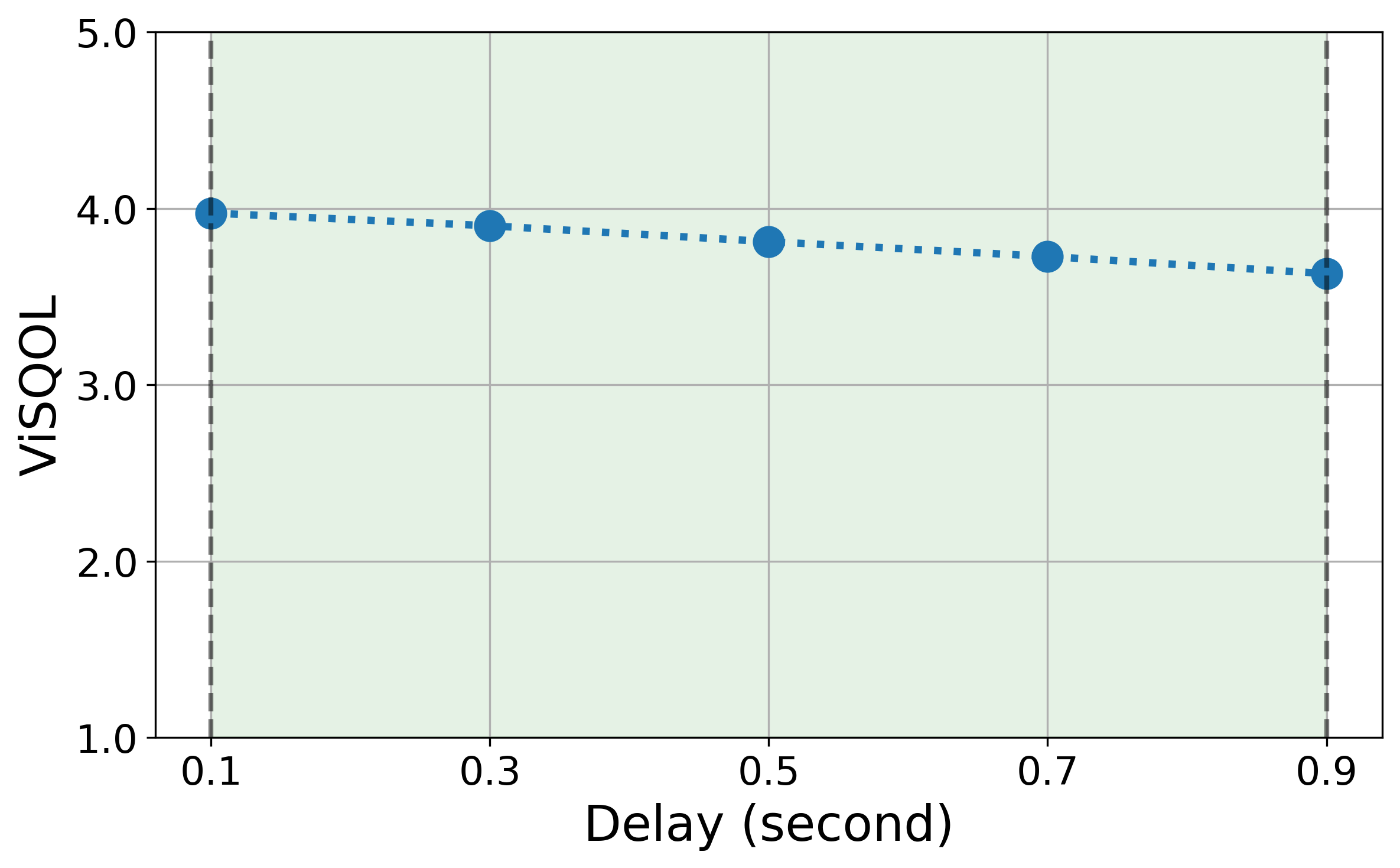}
        \caption{Echo}
        \label{fig:visqol_echo}
    \end{subfigure}

    \begin{subfigure}[b]{0.28\textwidth}
        \centering
        \includegraphics[width=\textwidth]{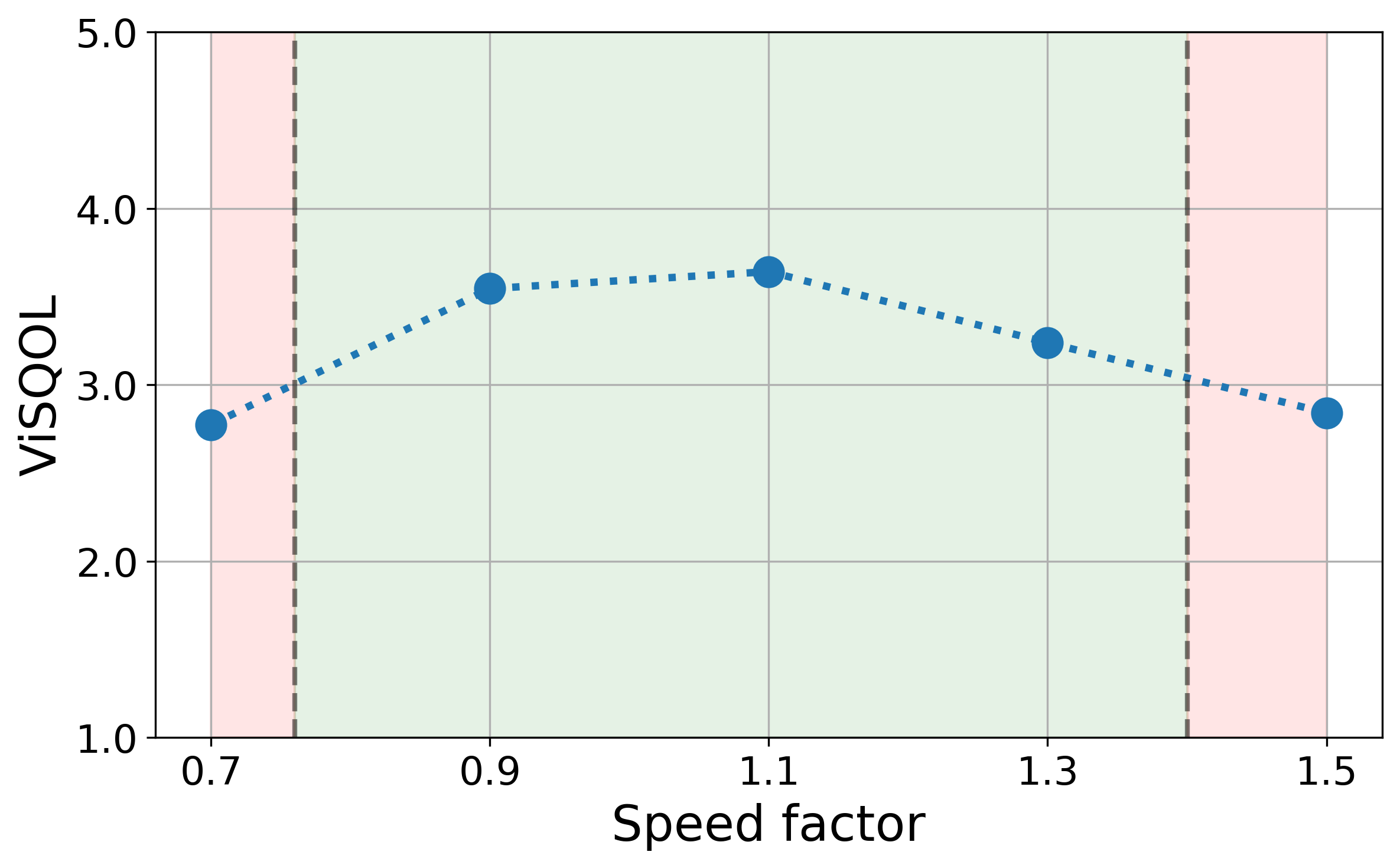}
        \caption{Time stretch}
        \label{fig:visqol_time_stretch}
    \end{subfigure}
    \begin{subfigure}[b]{0.28\textwidth}
        \centering
        \includegraphics[width=\textwidth]{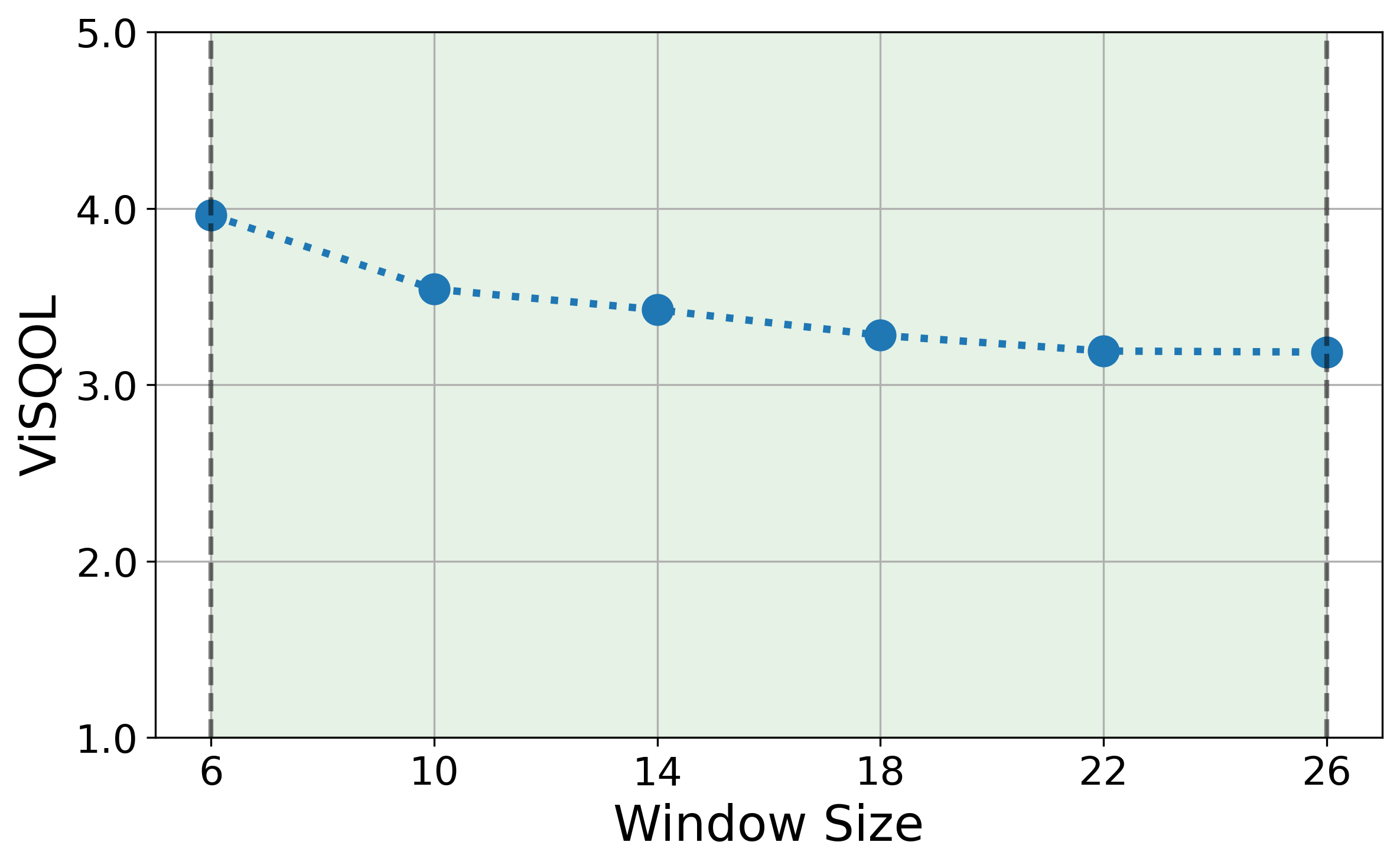}
        \caption{Smooth}
        \label{fig:visqol_smooth}
    \end{subfigure}
        \begin{subfigure}[b]{0.28\textwidth}
        \centering
        \includegraphics[width=\textwidth]{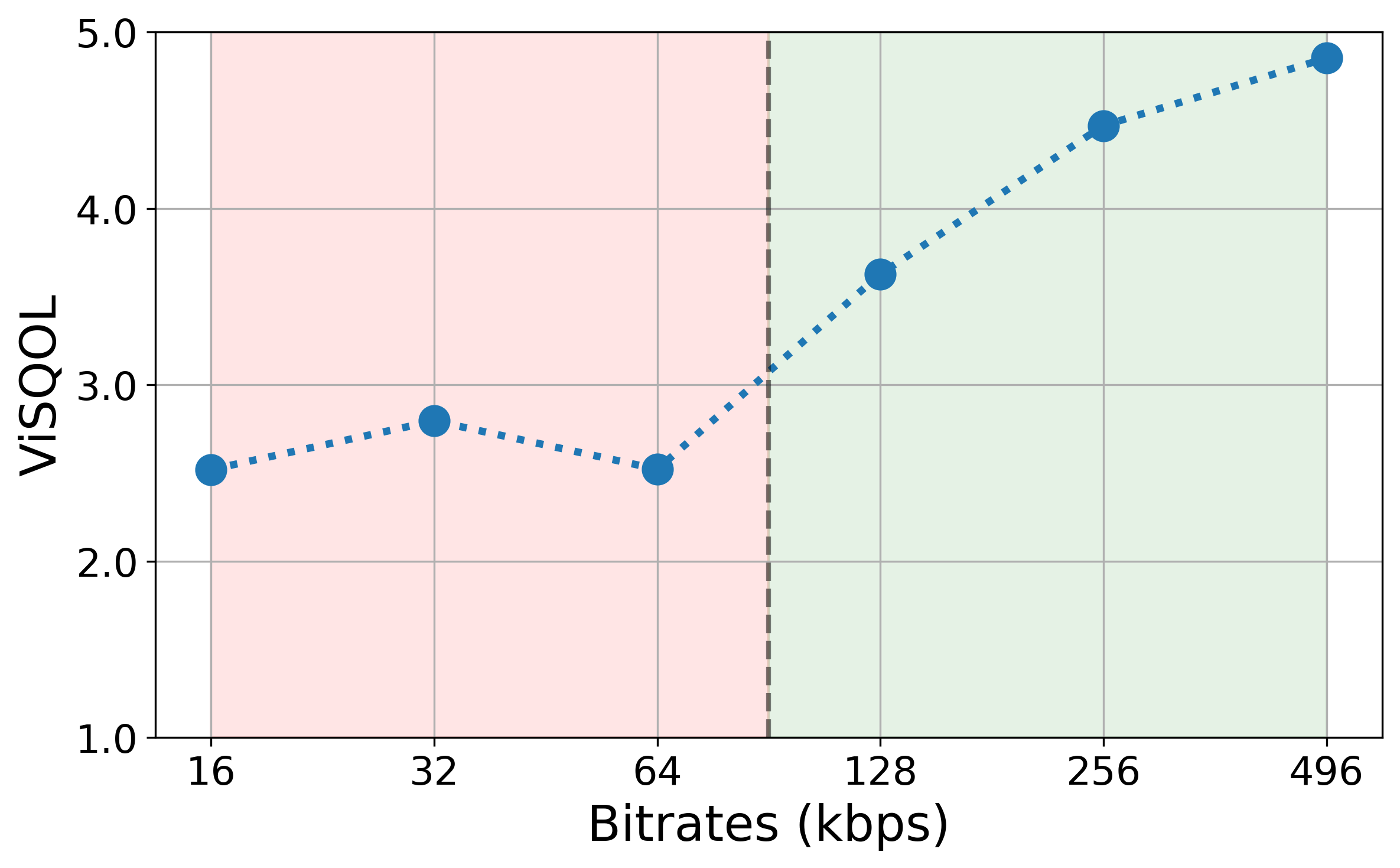}
        \caption{Opus}
        \label{fig:visqol_opus}
    \end{subfigure}

    \begin{subfigure}[b]{0.28\textwidth}
        \centering
        \includegraphics[width=\textwidth]{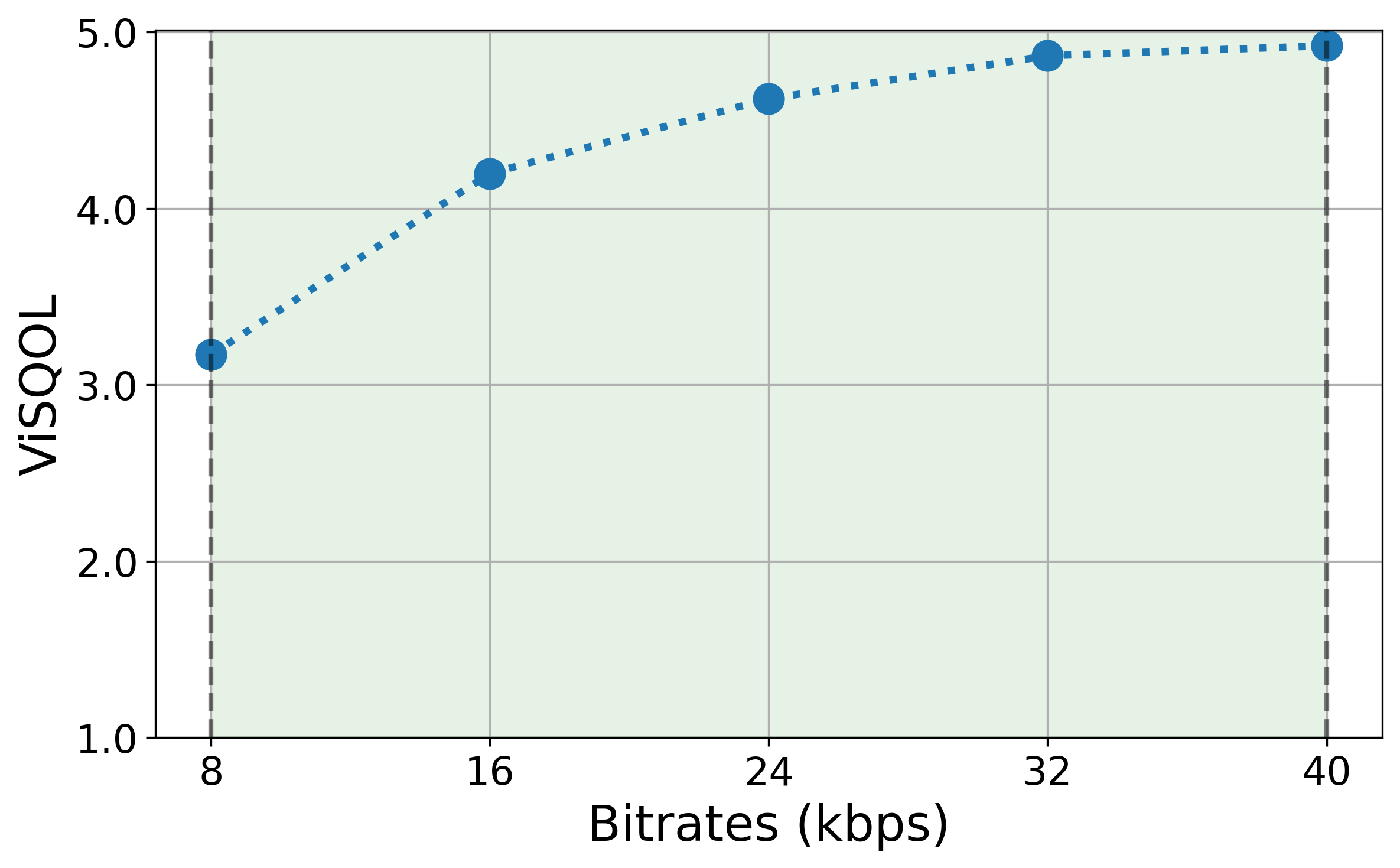}
        \caption{MP3}
        \label{fig:visqol_mp3}
    \end{subfigure}
    \begin{subfigure}[b]{0.28\textwidth}
        \centering
        \includegraphics[width=\textwidth]{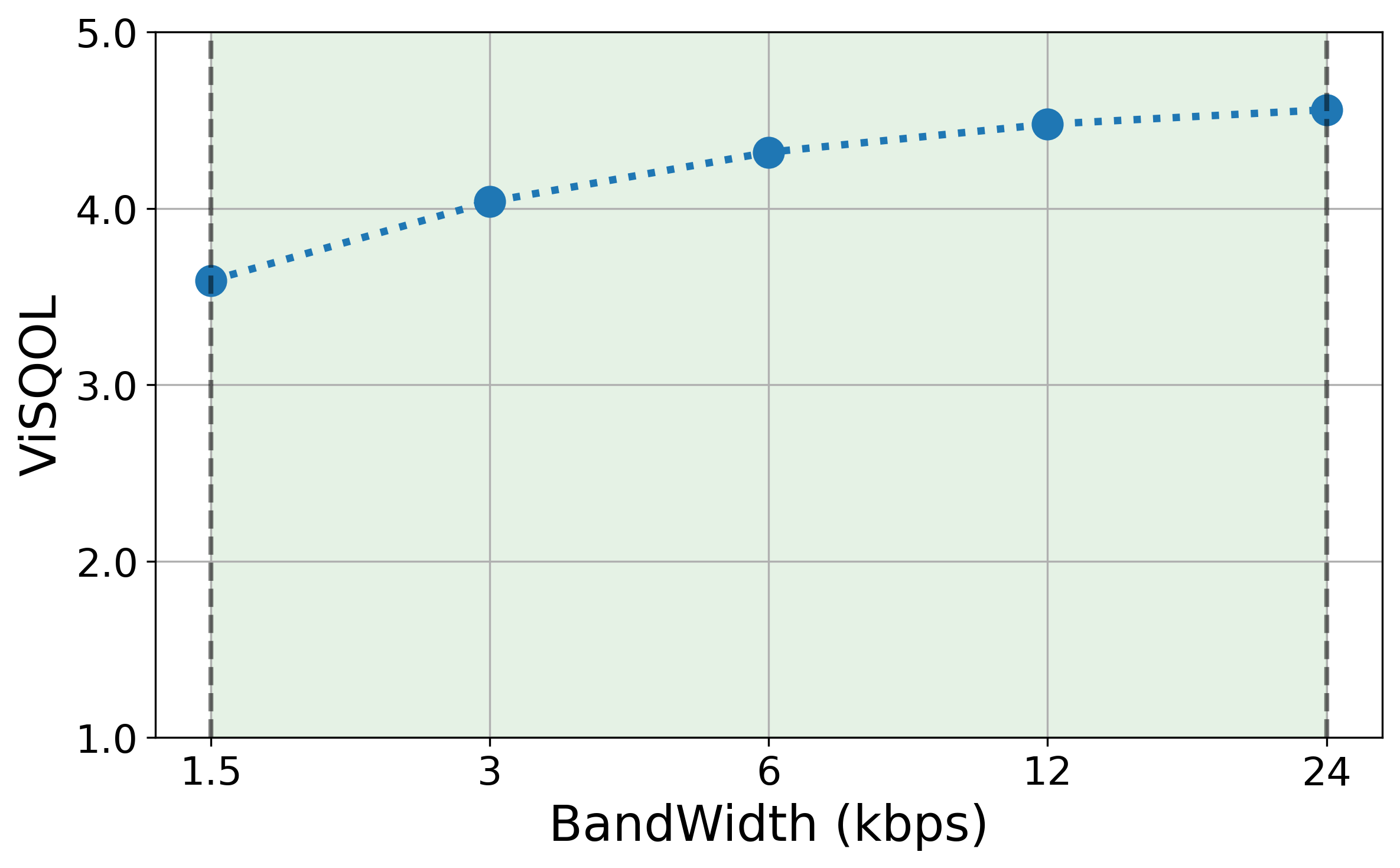}
        \caption{Encodec}
        \label{fig:visqol_encodec}
    \end{subfigure}
    \begin{subfigure}[b]{0.28\textwidth}
        \centering
        \includegraphics[width=\textwidth]{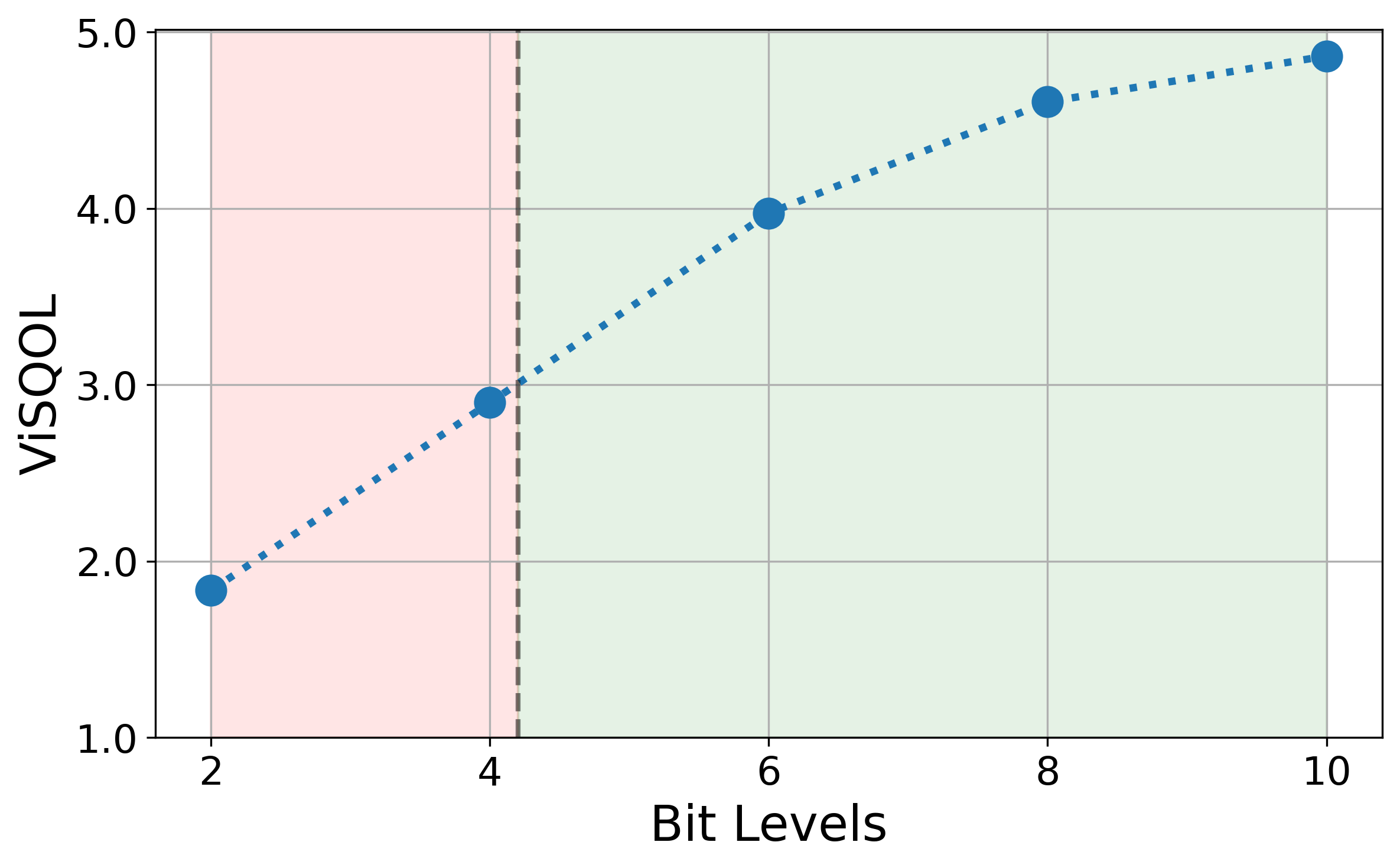}
        \caption{Quantization}
        \label{fig:visqol_quantization}
    \end{subfigure}

    \begin{subfigure}[b]{0.28\textwidth}
        \centering
        \includegraphics[width=\textwidth]{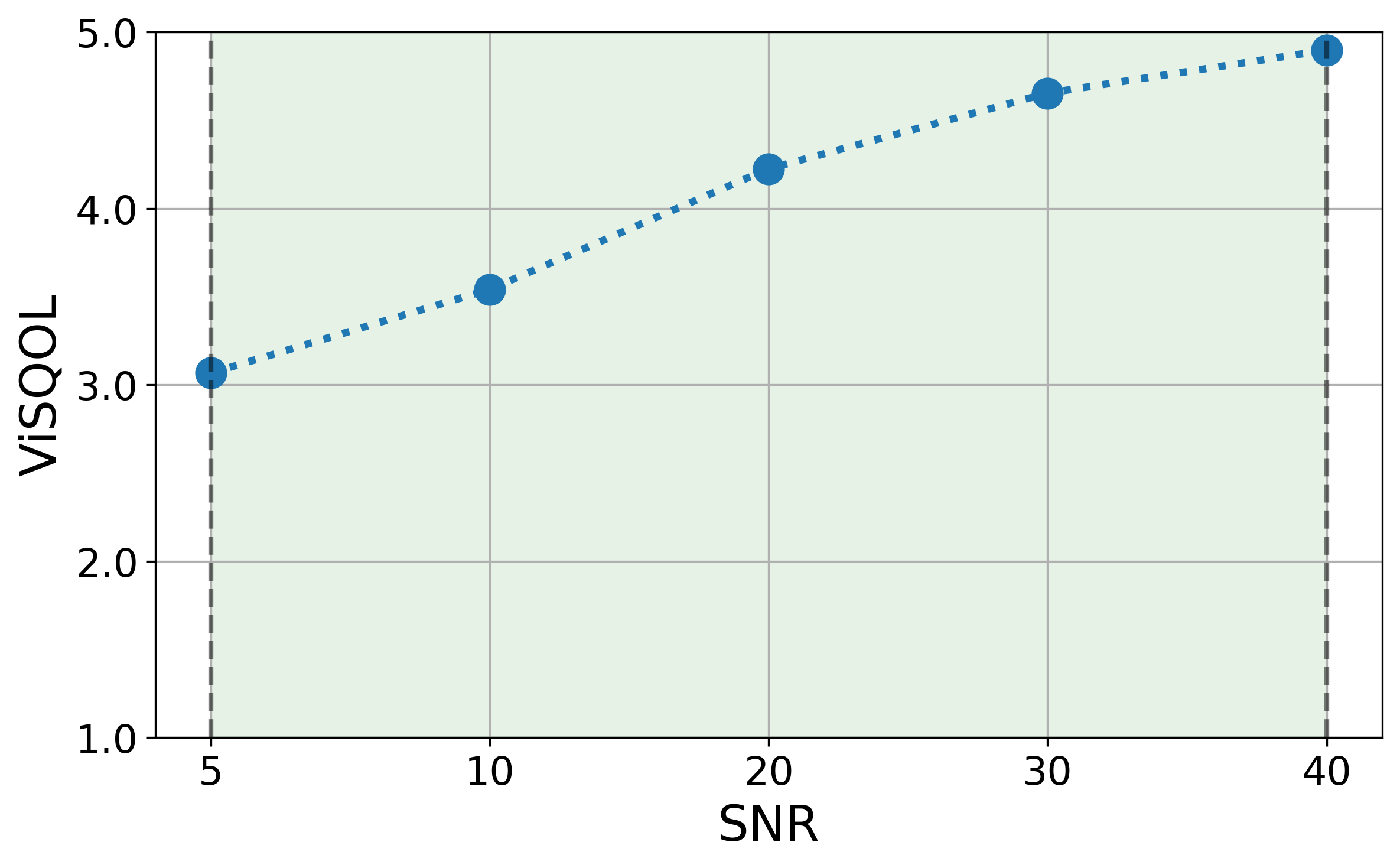}
        \caption{Background Music}
        \label{fig:visqol_music}
    \end{subfigure}
    \begin{subfigure}[b]{0.28\textwidth}
        \centering
        \includegraphics[width=\textwidth]{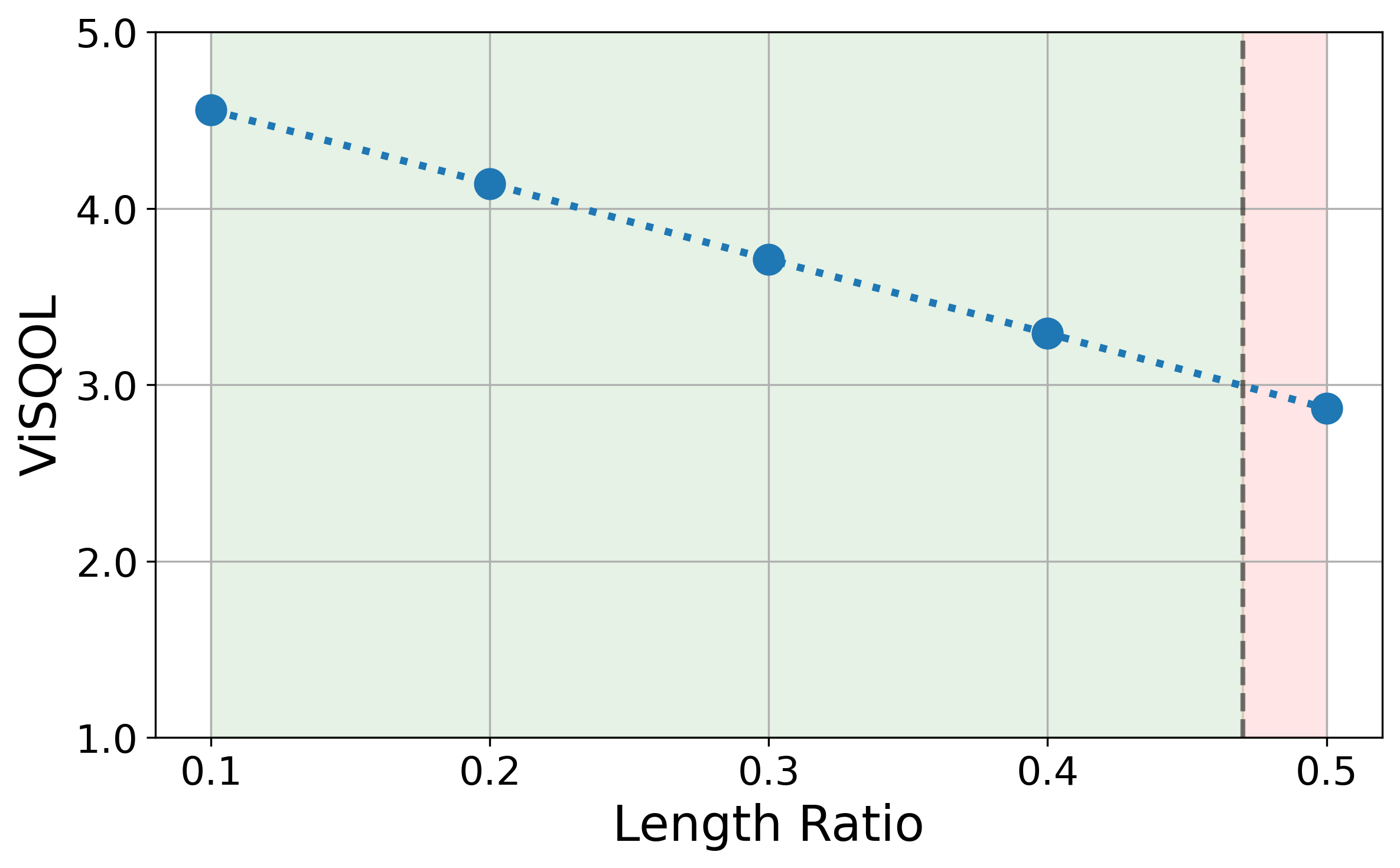}
        \caption{Silence Insertion}
        \label{fig:visqol_silence}
    \end{subfigure}
    \begin{subfigure}[b]{0.28\textwidth}
        \centering
        \includegraphics[width=\textwidth]{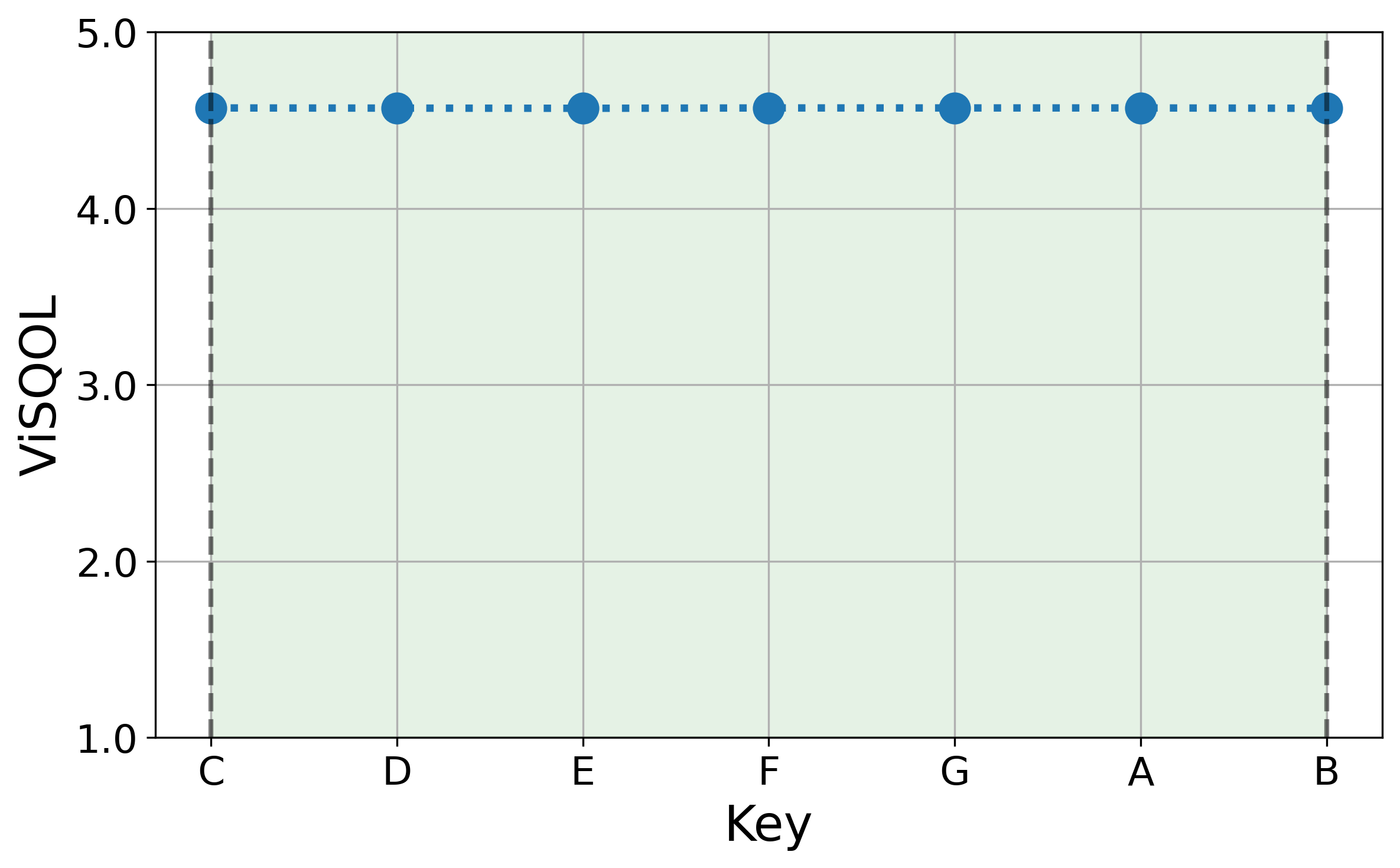}
        \caption{Autotune}
        \label{fig:visqol_autotune}
    \end{subfigure}

\caption{ViSQOL scores for various corruption types at different severity levels.}
\label{fig:audio_quality}
\end{figure*}

\textbf{Figure~\ref{fig:audio_quality}} illustrates the ViSQOL scores for various corruption types at different severity levels. A ViSQOL $\ge 3$ indicates acceptable audio quality (green region), while scores below this threshold indicate poor quality (red region). To better reflect real-world scenarios, we focus on evaluating models' robustness when the audio quality remains acceptable.

\section{Implementation details}

 We train AASIST, RawNet2, and RawGAT-ST with a learning rate of 0.0001, and LCNN and Spec.+ResNet with a learning rate of 0.0003 using Adam optimizer with momentum parameters $\beta_1 = 0.9$ and $\beta_2 = 0.999$, and a weight decay of $1 \times 10^{-4}$. The model is trained for 40 epochs without warmup, and we adopt a cosine decay learning rate scheduler. The batch size for AASIST, RawNet2, RawGAT-ST, LCNN, and Spec.+ResNet are 64,256, 32, 512, and 256, respectively.

For the foundation models, two linear layers are added after the encoder’s output, with the hidden layer dimension matching the dimension of the encoder’s output. We fine-tune all foundation models on the Wavefake training dataset for 3 epochs using the Adam optimizer with a learning rate of 0.00001 and a weight decay of 0.0005.

\section{Effectiveness of Speech Enhancement across different models.}
\label{appendix:speech_enhance}

\begin{figure*}[htbp]
    \centering
    \begin{subfigure}[b]{0.28\textwidth}
        \centering
        \includegraphics[width=\textwidth]{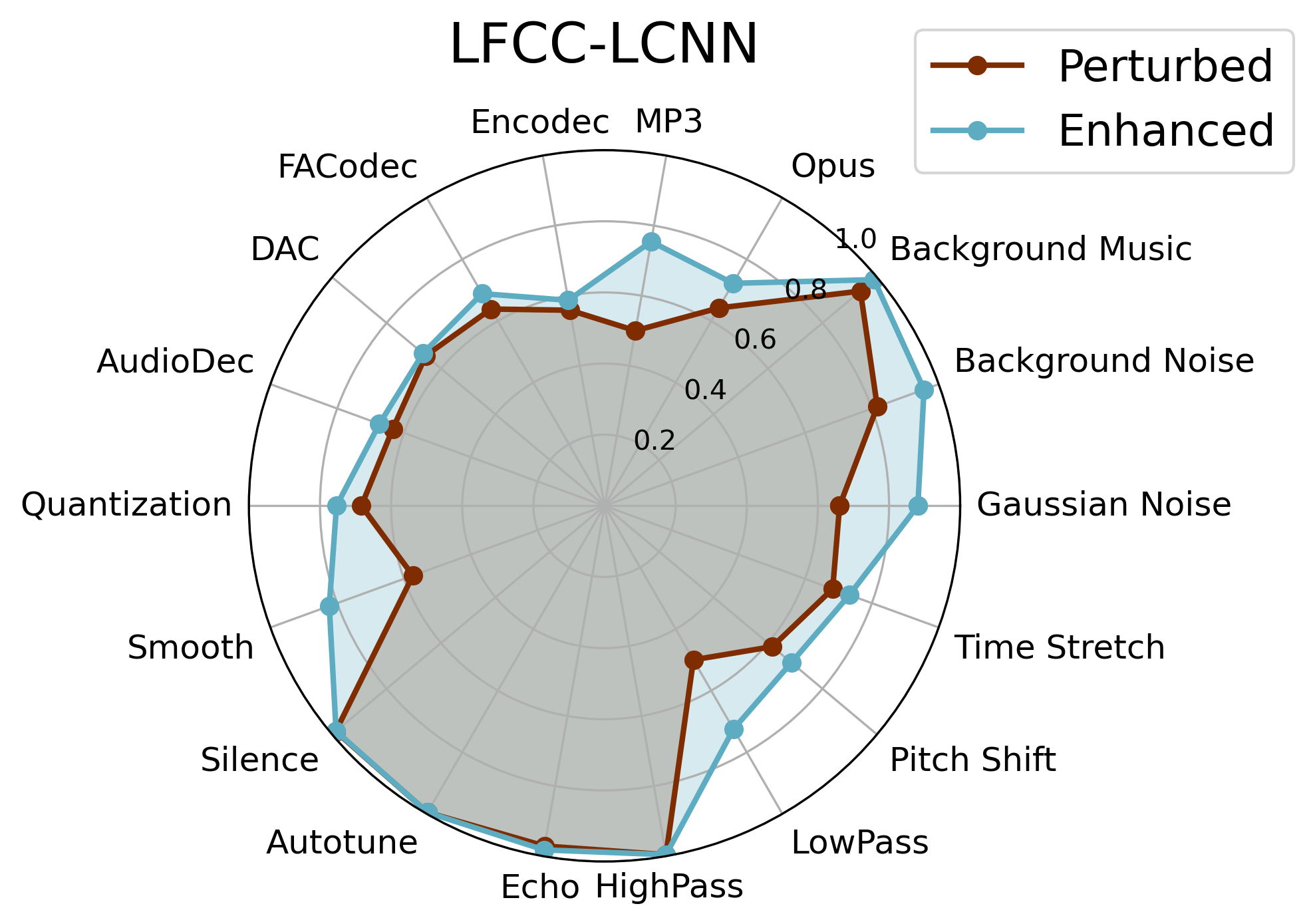}
        \caption{LFCC-LCNN}
        \label{fig:speech_enhance_lcnn}
    \end{subfigure}
    \begin{subfigure}[b]{0.28\textwidth}
        \centering
        \includegraphics[width=\textwidth]{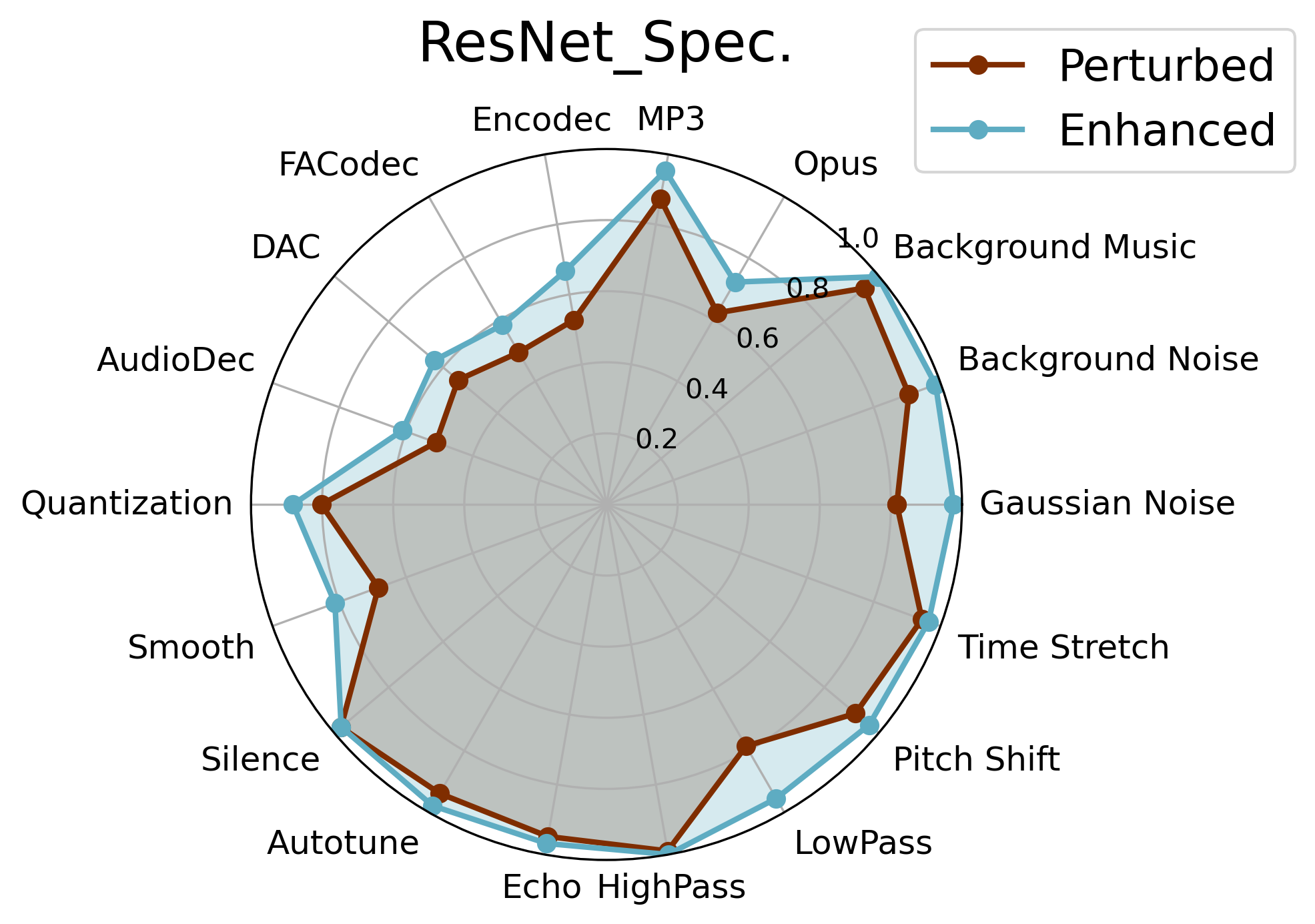}
        \caption{ResNet\_Spec.}
        \label{fig:speech_enhance_resnet}
    \end{subfigure}
    \begin{subfigure}[b]{0.28\textwidth}
        \centering
        \includegraphics[width=\textwidth]{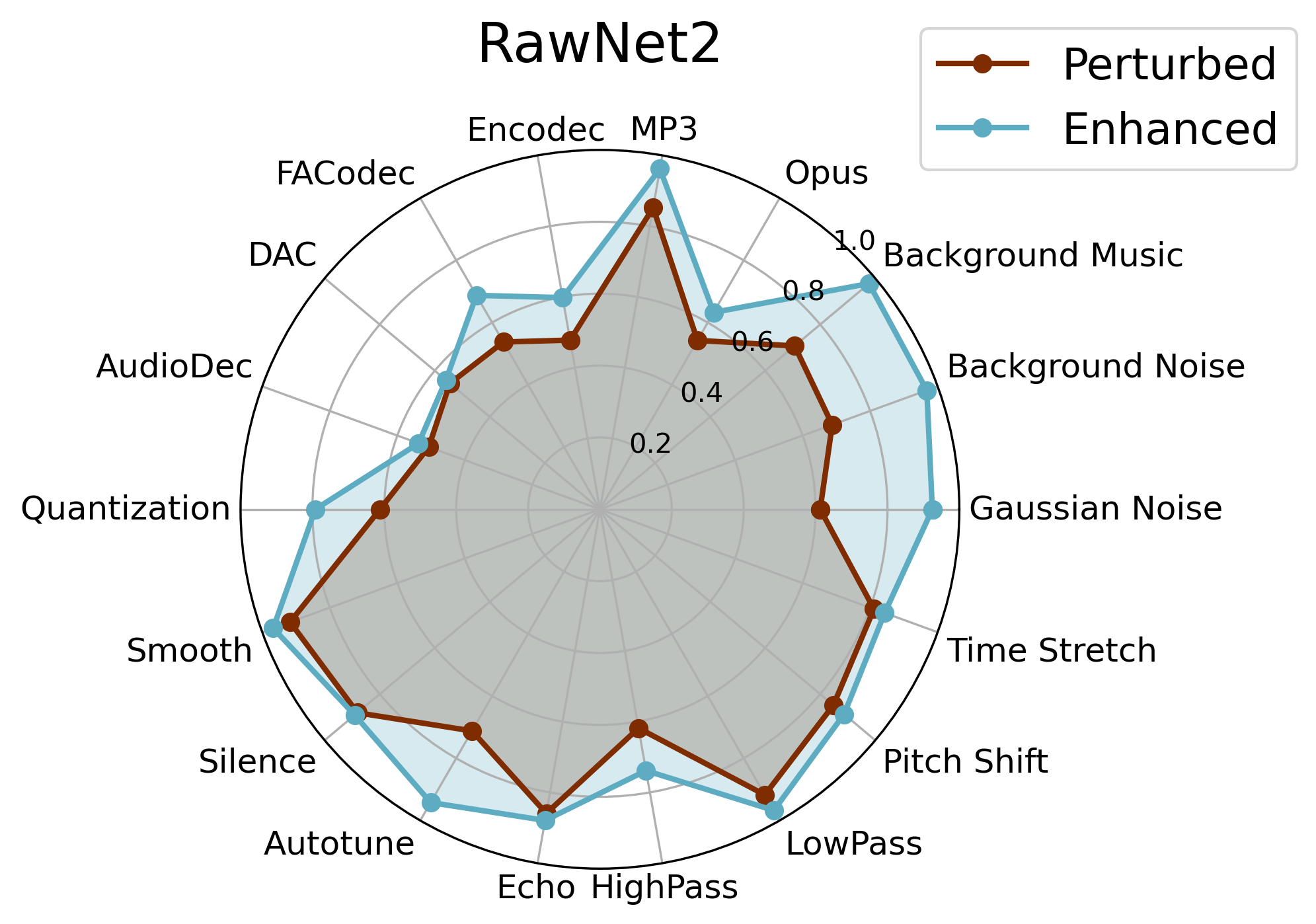}
        \caption{RawNet2}
        \label{fig:speech_enhance_rawnet2}
    \end{subfigure}
    
    \begin{subfigure}[b]{0.28\textwidth}
        \centering
        \includegraphics[width=\textwidth]{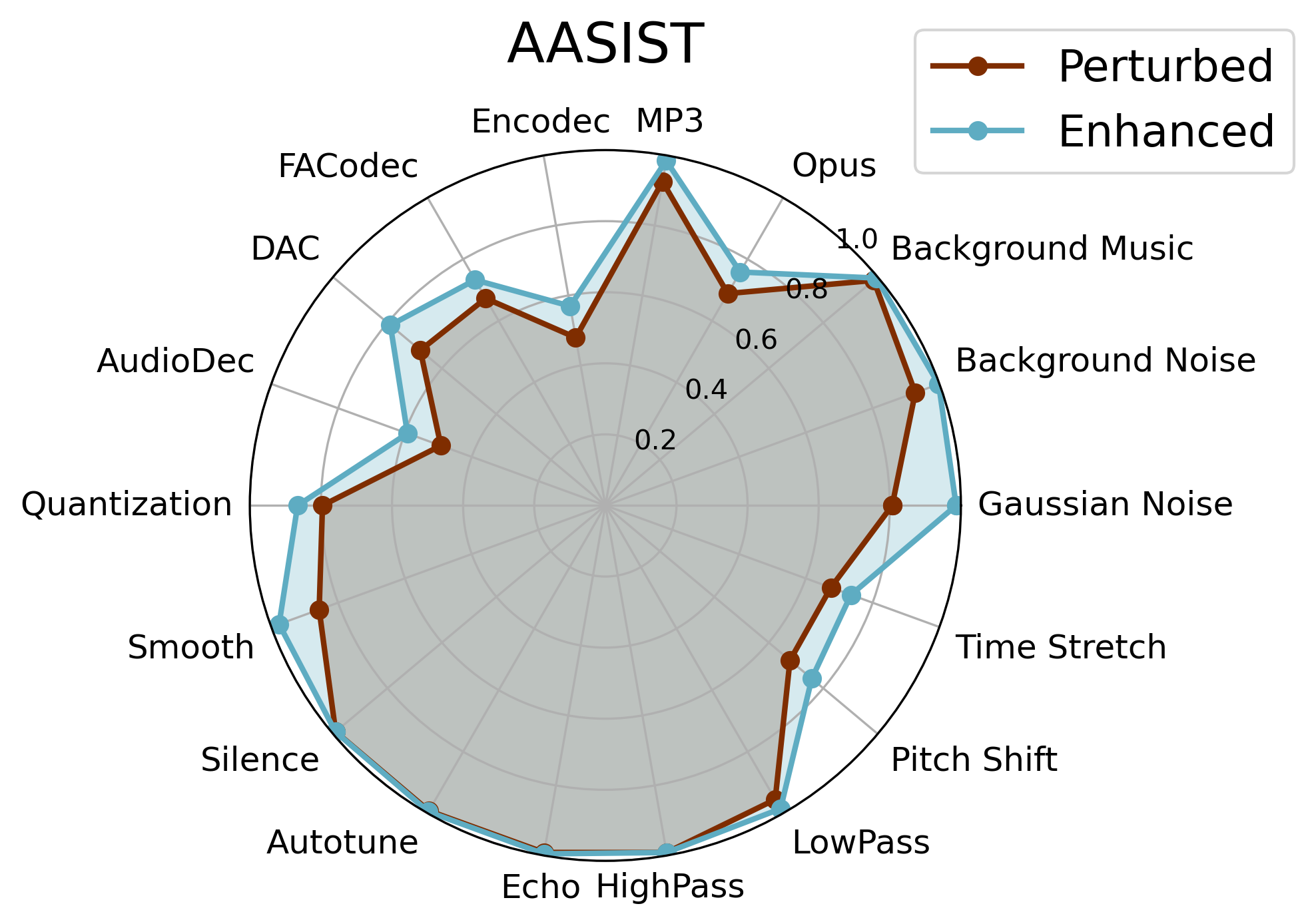}
        \caption{AASIST}
        \label{fig:speech_enhance_aasist}
    \end{subfigure}
    \begin{subfigure}[b]{0.28\textwidth}
        \centering
        \includegraphics[width=\textwidth]{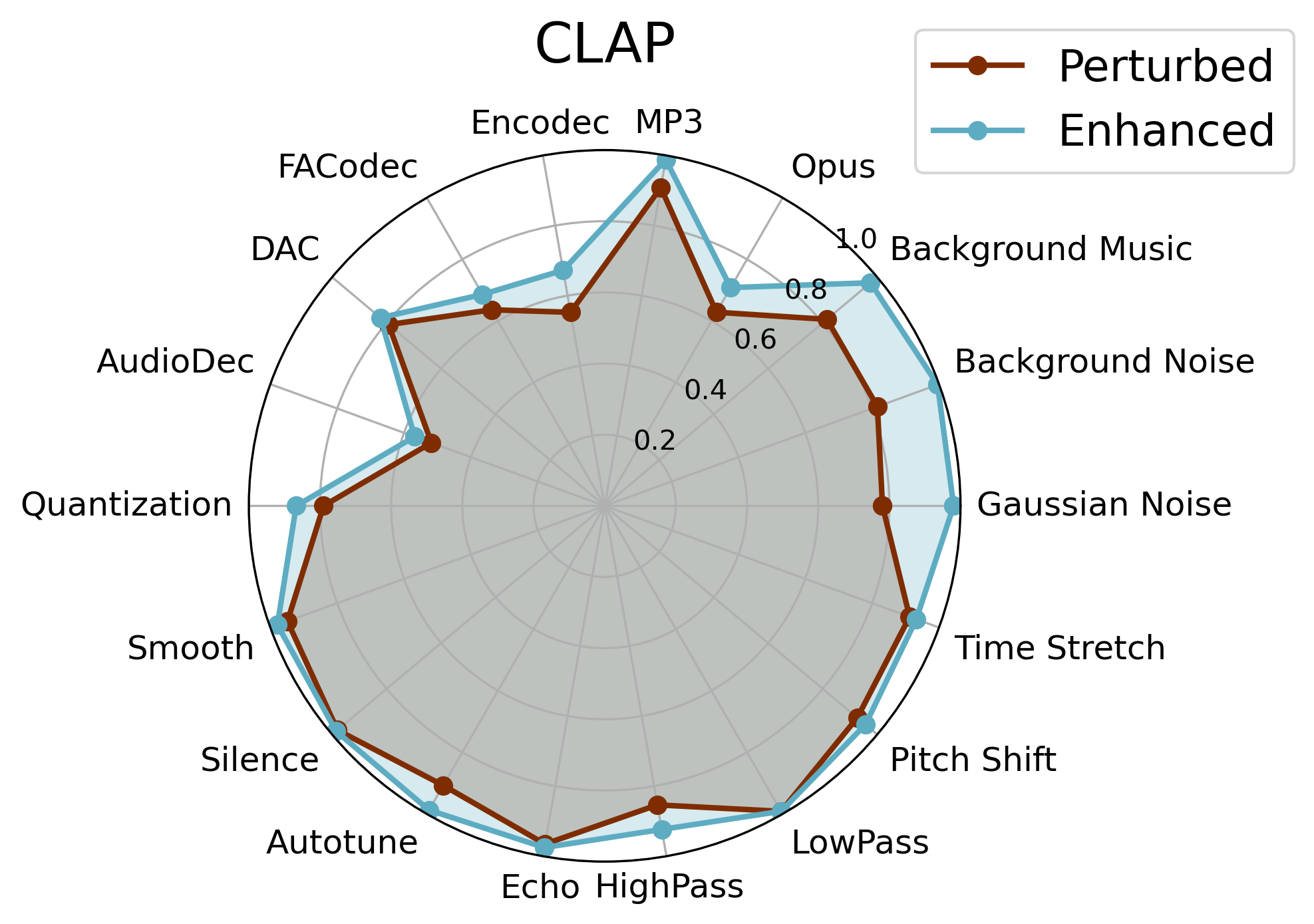}
        \caption{CLAP}
        \label{fig:speech_enhance_clap}
    \end{subfigure}
    \begin{subfigure}[b]{0.28\textwidth}
        \centering
        \includegraphics[width=\textwidth]{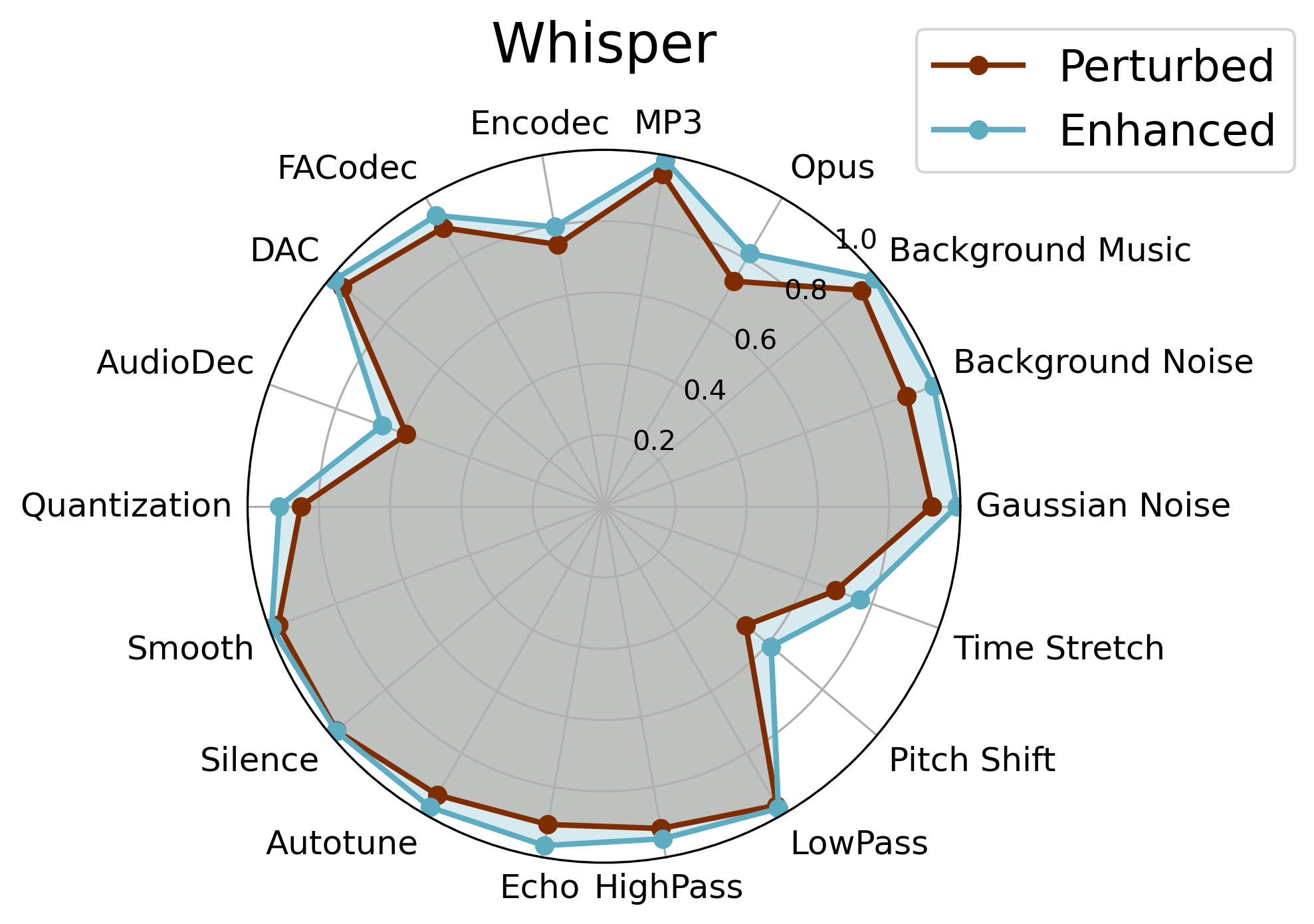}
        \caption{Whisper}
        \label{fig:speech_enhance_whisper}
    \end{subfigure}

    \begin{subfigure}[b]{0.28\textwidth}
        \centering
        \includegraphics[width=\textwidth]{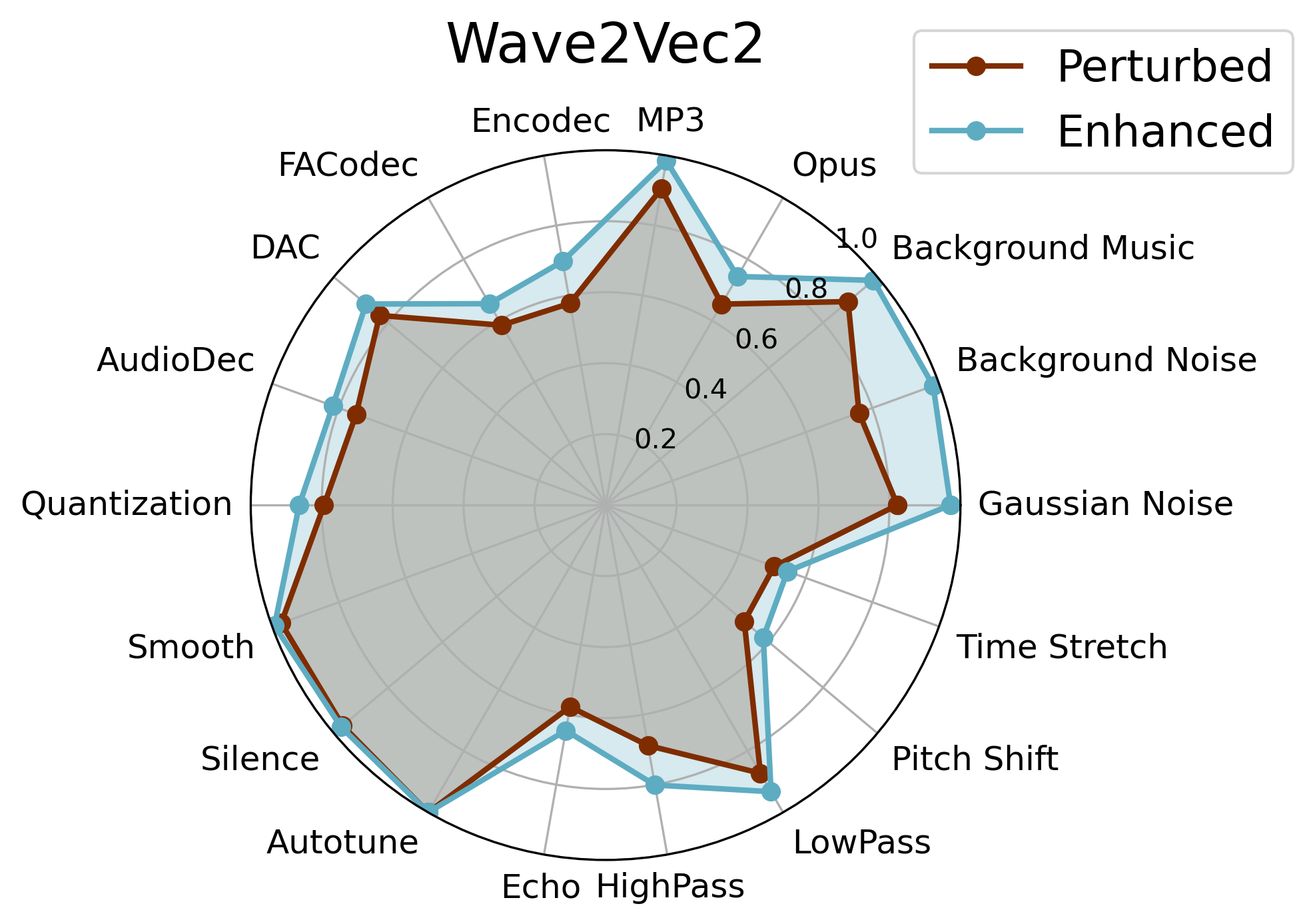}
        \caption{Wave2Vec2}
        \label{fig:speech_enhance_wave2vec2}
    \end{subfigure}
    \begin{subfigure}[b]{0.28\textwidth}
        \centering
        \includegraphics[width=\textwidth]{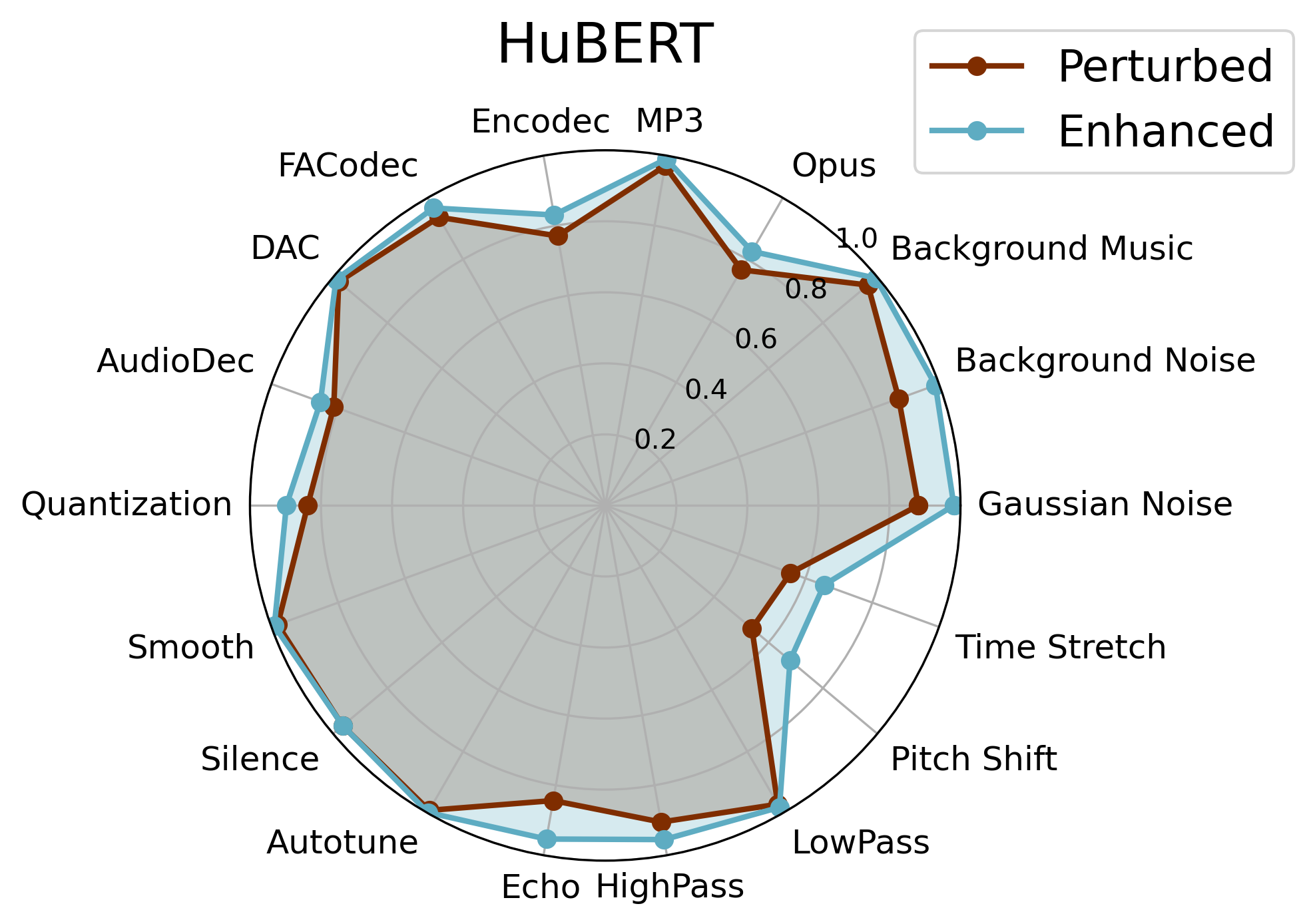}
        \caption{HuBERT}
        \label{fig:speech_enhance_hubert}
    \end{subfigure}
    \begin{subfigure}[b]{0.28\textwidth}
        \centering
        \includegraphics[width=\textwidth]{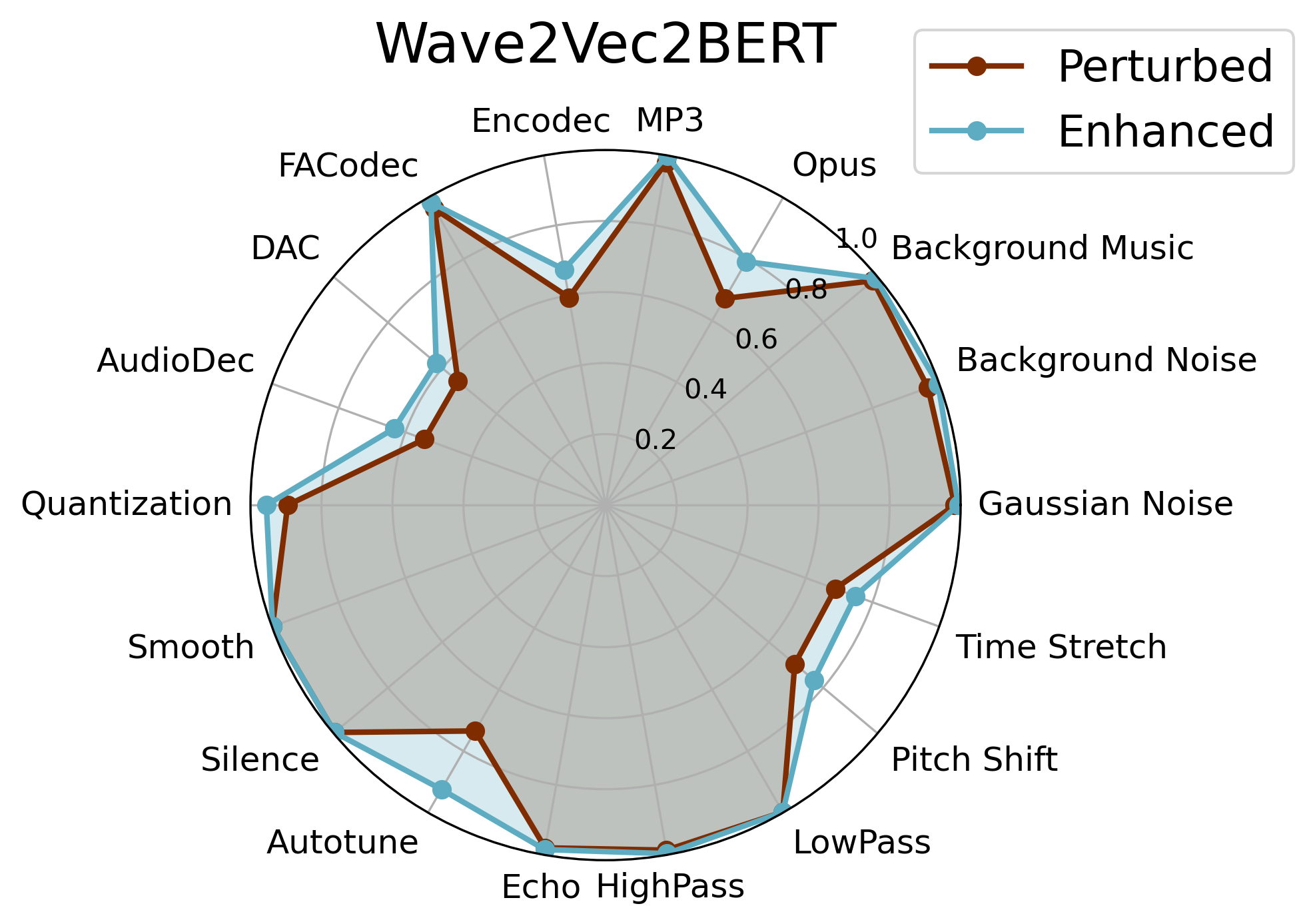}
        \caption{Wave2Vec2BERT}
        \label{fig:speech_enhance_wave2vec2bert}
    \end{subfigure}

\caption{Results across all perturbation types and models after applying speech enhancement.}
\label{fig:speech_enhancement}

\end{figure*}

\textbf{Figure~\ref{fig:speech_enhancement}} presents the detection accuracy of all evaluated models across different perturbation types, with results averaged over all severity levels. Consistent with our earlier discussion in the main script, the results show that speech enhancement improves robustness, most notably against noise perturbations. While improvements for modification and compression types are less significant, these findings suggest that speech enhancement could serve as a potential pre-processing defense to mitigate the adverse effects introduced by common real-world audio corruptions.

\newpage

\end{document}